\documentclass[11pt]{article}
\usepackage{fancyhdr,psfig,lastpage,latexsym}
\begin{document}
\newcommand{\fn}[1]{\footnote{ #1}}
\renewcommand{\textfraction}{0.05}
\renewcommand{\topfraction}{0.95}
\renewcommand{\bottomfraction}{0.95}
\renewcommand{\floatpagefraction}{0.95}
\newcommand{\mean}[1]{\left\langle #1 \right\rangle}
\newcommand{\tet}{\vartheta}
\newcommand{\eps}{\varepsilon}
\newcommand{\eqn}[1]{eq. (\ref{#1})}
\newcommand{\pic}[1]{Fig. \ref{#1}}
\newcommand{\name}[1]{{\rm #1}}
\newcommand{\bib}[4]{\item {#2} (#4). {#3}.}
\renewcommand{\cite}[1]{(#1)}
\newcommand{\et}{{\it et al.}}
\newcommand{\ul}[1]{\underline{#1}}
\newcommand{\ol}[1]{\overline{#1}}
\newcommand{\D}{\displaystyle}
\newcommand{\T}{\textstyle}
\newcommand{\SC}{\scriptstyle}
\newcommand{\SSC}{\scriptscriptstyle}


\vspace*{0.8cm}
\begin{center}
  {\Large \bf Modelling Migration and Economic Agglomeration \\[2mm]
    with Active Brownian Particles}\\[10mm]

{\large Frank Schweitzer}\\[4mm]
{\it Institute of Physics, Humboldt University,\\ 
Unter den Linden 6, 10099 Berlin, Germany}\\[2mm]
e-mail: frank@physik.hu-berlin.de\\[12mm]
\end{center}


\begin{abstract}
We propose a stochastic dynamic model of migration
  and economic aggregation in a system of employed (immobile) and
  unemployed (mobile) agents which respond to local wage gradients.
  Dependent on the local economic situation, described by a production
  function which includes cooperative effects, employed agents can become
  unemployed and vice versa. The spatio-temporal distribution of employed
  and unemployed agents is investigated both analytically and by means of
  stochastic computer simulations. We find the establishment of distinct
  economic centers out of a random initial distribution. The evolution of
  these centers occurs in two different stages: (i) small economic
  centers are formed based on the positive feedback of mutual
  stimulation/cooperation among the agents, (ii) some of the small
  centers grow at the expense of others, which finally leads to the
  concentration of the labor force in different extended economic
  regions. This crossover to large-scale production is accompanied by an
  increase in the unemployment rate. We observe a stable coexistence
  between these regions, although they exist in an internal
  quasistationary non-equilibrium state and still follow a stochastic
  eigendynamics.

\end{abstract}

\section{Introduction}
In recent years, there has been an increasing interest to link the
discussion about complex phenomena in natural science, such as physics or
biology, in particular to those in life science, such as sociology,
economy or regional planning. One of the challenges is to reveal
cross-links between the dynamic models used in the specific fields, in
order to find out pieces for a common theory of self-organization and
evolution of complexity. 

With respect to economic and urban problems, this perspective seems
rather new. In his essay about self-organization in economics, Krugman
(1996b) states: ``so far this movement has largely passed economics by.''
However, recent years have also seen numerous attempts for developing a
new, evolutionary view of economics \cite{Anderson \et, 1988, Wei--Bin,
  1991, Arthur \et, 1997, Silverberg, 1997, Schweitzer and Silverberg,
  1998}

In a world of fast growing population and intercontinental trade
relations, the \emph{local} emergence of new economic centers on one
hand, and the \emph{global} competition of existing economic centers on
the other hand has a major impact on the socio-economic and political
stability of our future world. This problem also confronts the natural
sciences on their way to a new understanding of complex phenomena. This
paper aims to contribute to this discussion by adapting a model of
interactive structure formation, which has already proved its versatility
in a variety of applications, to the problem of economic agglomeration.

Spatio-temporal pattern formation in urban and economic systems has been
investigated for quite a long time \cite{Henderson, 1988, Fujita, 1989,
  Puu, 1993, Krugman 1991, 1996a}. There is also a well known German
tradition in location theory, associated with names like von Th\"unen,
Weber, 
Christaller (1933) and L\"osch (1940).  Recent
approaches (sparsely recognized especially by American economists)
 tackle the problem of settlement formation, location theory
and city size distribution within a stochastic dynamical theory, based on
master equations and economically motivated utility functions \cite{
Dendrinos and Haag, 1984, 
Weidlich and Munz, 1990 a,b, Haag \et, 1992, 1994, Weidlich, 1991, 
1997}.

Different from these investigations, the current paper focuses on an
agent based dynamics of economic concentration. Agent models, which have
originally been developed in the Artificial Life community \cite{Maes, 
1990, 
Steels, 1995} 
recently turned out to be a suitable
tool for describing economic interaction \cite{
Anderson \et, 1988, Marimon \et, 1990, 
Holland and Miller, 1991, Lane, 1992, Arthur, 1993, Kirman, 1993, 
Epstein and Axtell, 1996, Arthur \et, 1997}.

The rational agent model, one of the standard paradigms of neoclassical
economic theroy \cite{Silverberg and Verspagen, 1994}, is based, among
others, on the assumption of the agent's complete knowledge of all
possible actions and their outcomes or a known probability distribution
over outcomes, and the common knowledge assumption, i.e. that the agent
knows that all other agents know exactly what he knows and are equally
rational.

In this particular form, the rational agent is just one example of a
\emph{complex agent}, which can be regarded as an autonomous entity with
either knowledge based or behavior based rules \cite{Maes, 1990},
performing complex actions, such as BDI (belief-desire-interactions)
\cite{M\"uller \et, 1997}.  The complex agent i.e. is capable of
specialization, learning, genetic evolution, etc.  However, the freedom
to define rules and interactions for the agents, could very soon turn out
to be a pitfall, because of the combinatoric explosion of the state
space: for 1000 Agents with 10 rules, the state space contains
$10^{1000}$ possibilities, hence, almost every desirable result could be
produced from such a simulation model.

The alternative to the complex agent, promoted in this paper, could be
the \emph{minimalistic agent}, which acts on the possible simplest set of
rules, without deliberative actions.  Instead of specialization, the
minimalistic agent model is based on a large number of ``identical''
agents, and the focus is mainly on cooperative interaction instead of
autonomous action. 

A version of the minimalistic agent model, which allows to apply the
advanced methods developed in statistical physics and stochastic theory,
is based on active Brownian particles \cite{Steuernagel \et, 1994,
  Schimansky-Geier \et, 1995, 1997, Schweitzer, 1997, Ebeling \et, 1998}.
Generally, active Brownian particles are Brownian particles with internal
degrees of freedom.
As a specific action,  the active Brownian particles (or active
walkers, within a discrete approximation) are able  to generate
a self-consistent field which in turn influences their further movement
and behavior.  This non-linear feedback between the particles and the
field generated by themselves results in an \emph{interactive structure
  formation} process on the macroscopic level. Hence, these models have
been used to simulate a broad variety of pattern formations in complex
systems, ranging from physical to biological and social systems (
Lam and Pochy, 1993, Schweitzer and Schimansky-Geier, 1994, Ben-Jacob
\et, 1995, Lam, 1995, 
Schweitzer \et, 1997, 
Helbing \et, 1997, Schweitzer and Steinbrink, 1997).

In the following, the active Brownian particles are regarded as economic
agents with two different internal states and a very simple behavior:
employed agents, which are immobile, generate a wage field (as the result
of their work), while unemployed agents migrate guided by local gradients
in the wage field, i.e. they try to move toward those regions in their
vicinity with a high productivity and a high value of the wage field.
Further unemployed agents can be employed, dependent on the local
economic situation, but employed agents also can be unemployed.

In Sect. 2 we present our stochastic approach to the problem of
employment and migration. In Sect. 3, the general model is used to derive
a special dynamics recently applied by Krugman (1992, 1996b) for economic
aggregation. In Sect. 4, we present the economic assumptions involved in
our model, while Sect. 5 describes the results of computer simulations,
which show economic aggregation on two time scales. Sect. 6 gives some
conclusions and an outlook of possible extension of the model presented. 

\section{Dynamic Model of Migration and Employment}

Let us consider a two-dimensional system with $i=1,...,N$ economic
agents, which are represented by active Brownian particles. These
particles are characterized by two variables: their current location,
given by the space coordinate $r_{i}$, and an internal state,
$\theta_{i}$, which could be either one or zero: $\theta \in \{0,1\}$.
Active particles with the internal state $\theta=0$ are considered
employed agents, $C_{0}$, active particles with $\theta=1$ are considered
unemployed agents, $C_{1}$.  With a certain rate, $k^{+}$ (hiring rate),
an unemployed agent becomes employed, while with a rate $k^{-}$ (firing
rate) an employed agent becomes unemployed, which can be expressed by the
symbolic reaction:
\begin{equation}
  \label{react}
  C_{0} \;\; {{k^{-} \atop -\!\!\!-\!\!\!-\!\!\!\longrightarrow} 
\atop {\longleftarrow\!\!\!-\!\!\!-\!\!\!-  \atop  k^{+}}} \;\; C_{1} 
\end{equation}
Employed agents are considered immobile, while unemployed agents are able
to migrate.  The movement of a migrant may depend both on erratic
circumstances and on deterministic forces which attract him to a certain
place.  Within a stochastic approach, this movement can be described by
the following overdamped Langevin equation:
\begin{equation}
\frac{dr_i}{dt}= f(r_{i})
+ \sqrt{2 \eps}\;\xi_{i}(t)
\label{langev-red}
\end{equation} 
$f(r_{i})$ describes the local value of a deterministic force, which
influences the motion of the agent. We note here, that the agent is
\emph{not} subject to long-range forces, which may attract him over large
distances, but only to a \emph{local} force. That means the migrant
does \emph{not} count on global information which guide his movement, but
responds only to local information, which will be specified later. The
second term in \eqn{langev-red} describes random influences on the
movement of the individuals modelled by a stochastic force $\xi(t)$,
which is assumed to be Gaussian white noise:
\begin{equation}
\mean{\xi_{i}(t)} = 0 \quad \mbox{;}\quad 
\mean{ \xi_{i}(t) \xi_{j}(t')} = \delta_{ij} \delta (t-t`).
\label{noise}
\end{equation}
$\eps$ is a measure of the strength of the stochastic force. As
\eqn{langev-red} indicates the unemployed agent will move in a very
predictable way if the guiding force is large and the stochastic
influence is small, and he will act rather randomly in the opposite case.

The current state of the agent community can be described by the
canonical $N$-particle distribution function
$P(\theta_{1},r_{1},...,\theta_{N},r_{N},t)$, which gives the probability
to find the $N$ agents with the internal states
$\theta_{1},...,\theta_{N}$ in the vicinity of $r_1,....,r_N$ at time
$t$. The change of this probability can be described by a multi-variate
master equation \cite{Feistel and Ebeling, 1989}, which considers both
changes in the internal states and migration of the agents. Here, we
restrict the discussion to the spatio-temporal densities
$n_{\theta}(r,t)$ of unemployed and employed agents, which can be
formally derived from the $N$-particle distribution function:
\begin{equation} 
n_{\theta}(r,t)=\sum_{i=1}^N \delta_{\theta,\theta_{i}}\;\int_{A} 
\delta(r-r_{i})\;P(\theta_{1},r_{1}...,\theta_{N},r_{N},t)\; 
dr_1 ... dr_{N} 
\label{dens} 
\end{equation} 
$\delta_{\theta,\theta_{i}}$ is the Kronecker Delta function for
discrete variables, which is $1$ only for $\theta=\theta_{i}$ and $0$
otherwise, while $\delta(r-r_{i})$ is Dirac's Delta function
used for continuous variables, and $A$ is the system size (surface area).
With $\theta \in \{0,1\}$, we obtain the spatio-temporal density of the
employed agents, $n_{0}(r,t)$, and of the unemployed agents,
$n_{1}(r,t)$. For simplicity, we omit the index $\theta$, by defining:
\begin{equation}
  \label{nl}
  n_{0}(r,t)=l(r,t)\;;\;\;n_{1}(r,t)=n(r,t)
\end{equation}
We only assume that the total
number of agents is constant, while the density of employed and unemployed
agents can change in space and time:
\begin{equation}
  \label{total}
  N_{0}(t)=\int\nolimits_{A} l(r,t)\,dr\;;\;\;
N_{1}(t)=\int\nolimits_{A} n(r,t)\,dr\;;\;\;
  N_{0}(t)+N_{1}(t)=N  
\end{equation}
The density of the employed agents can be changed only by local
``hiring'' and ``firing'' processes, which with respect to \eqn{react},
can be described by the reaction equation:
\begin{equation}
  \label{react2}
  \frac{\partial}{\partial t}l(r,t)= k^{+}\,n(r,t)\;-\;k^{-}\,l(r,t)
\end{equation}
The density of the unemployed agents can be changed by two processes, (i)
migration and (ii) hiring of unemployed and firing of employed
agents. With respect to \eqn{langev-red}, which describes the movement,
we can derive the  following Fokker-Planck equation:  
\begin{samepage}
\begin{eqnarray}
\label{fpe-t}
\frac{\partial}{\partial t}\,n(r,t)\; = \;  
&-& \frac{\partial}{\partial r} f(r,t)\; n(r,t)
+ D_n\;\frac{\partial^{2}}{\partial r^{2}} n(r,t) \nonumber \\
& -&  k^{+}\,n(r,t) + k^{-}\,l(r,t)
\end{eqnarray}
\end{samepage}
The first term of the r.h.s. of \eqn{fpe-t} describes the change of
the local density due to the force $f(r,t)$, the second term describes
the migration of the unemployed agents in terms of a diffusion process
with $D_{n}=\eps$ beeing the diffusion coefficient, the third and the
fourth term describe local changes of the density due to ``hiring'' or
``firing'' of agents.

By now, we have a complete dynamic model which describes  local changes
of employment and unemployment, as well as the migration of
unemployed agents. 
However, so far some of the important features of this dynamic model are
not specified, namely (i) the deterministic influences on a single
migrant, expressed by $f(r,t)$, (ii) the hiring and the firing rates,
$k^{+},\; k^{-}$, which locally change the employment density. These
variables depend of course on local economic conditions, hence we will
need additional economic assumptions.

\section{Krugman's ``Law of Motion of the Economy''}
Krugman (1992) (see also Krugman (1996b) for the results) discusses a
dynamic spatial model, where ``workers are assumed to move toward
locations that offer them higher real wages''.  Krugman makes no attempt
``to model the moving decisions explicitely'', but he assumes the
following ``law of motion of the economy'':
\begin{equation}
  \label{krugman}
  \frac{d \; \lambda_{j}}{dt} = \rho \;
  \lambda_{j}\;(\omega_{j}-\bar{\omega})
\end{equation}
Here, $\lambda_{j}$ is the ``share of the manufactoring labor force in
location $j$'' and $\omega_{j}$ is the real wage at location
$j$. $\bar{\omega}$ is the ``average real wage'', which is defined as:
\begin{equation}
  \label{mean-w}
  \bar{\omega}=\frac{\sum_{j} \lambda_{j}\; \omega_{j}}{\sum_{j} %
  \lambda_{j}}\;\;;\quad \sum_{j} \lambda_{j} = 1 
\end{equation}
Here, the sum goes over all different regions, where $j$ serves as a
space coordinate, and it is assumed that ``at any point in time there
will be location-by-location full employment''. The assumed law of motion
of the economy, \eqn{krugman}, then means that ``workers move away from
locations with below-average real wages and towards sites with
above-average real wages''.

In the present form, the assumed law of motion of the economy involves
some shortages:
\begin{enumerate}
\item A constant total number of employed workers is assumed, 
  a change of the total number or unemploment is not discussed.
\item  Employed workers are considered to move. In fact, if a worker
  wants to move to a place which offers him higher wages, he first has to
  be a free, that means an unemployed worker, then moves and then has to
  be reemployed at the new location, again. This process is completely
  neglected (or else, it is assumed that it occurs with infinite
  velocity). 
\item  Workers at location $j$ always exactly know about
  the average wage in the system. It is not explained where they get the
  information about the average wage from.
\item  Workers move immediately  if their
  wage is below the average, regardless of the migration distance to the
  places with higher wages, and with no doubt about their reemployment.
\end{enumerate}
In the following, Krugman's law of motion of the economy shall be derived
from the general dynamic model presented in Sect. 2., in order to
elucidate the implicit assumptions which lead to \eqn{krugman}. The
derivation is based on four approximations:

\textbf{(i)} In his paper, \name{Krugman (1992)} does not discuss
unemployment. But, with respect to the dynamic model presented in Sect.
2, the unemployed agents exist in an overwhelming large number. Thus,
the first assumption to derive Krugmans law of motion is, the local
change of the density of unemployed agents due to hiring and firing can
be simply neglected.

\textbf{(ii)} Further, it is assumed that the spatial distribution of
unemployed agents is in a \emph{quasistationary} state.  This does not
mean that the distribution does not change, but that the distribution
relaxes fast into a quasistationary equilibrium, compared to the
distribution of the employed agents.

With the assumptions (i) and (ii), \eqn{fpe-t} for the density of the
unemployed agents reduces to:
\begin{equation}
\label{fpe-st}
\frac{\partial}{\partial t}\,n(r,t)\; = \;  
- \frac{\partial}{\partial r} f(r,t)\;n(r,t)
+ D_n\;\frac{\partial^{2}}{\partial r^{2}} n(r,t) \;=\; 0
\end{equation}
Integration of \eqn{fpe-st} leads to the known canonical distribution: 
\begin{eqnarray}
  \label{stat}
  n^{stat}(r,t) &=&\bar{n} \frac{\exp \left[\int f(r,t)/D_{n}\; 
dr\right]}%
{\mean{\exp \left[\int f(r,t)/D_{n}\; dr \right]}}  \\
\mean{\exp \left[\int f(r,t)/D_{n}\; dr \right]} &=& \frac{1}{A} 
\int_{A} \exp \left[ \int f(r',t)/D_{n} \; dr'\right] \; dr \nonumber
\end{eqnarray}
where the expression $\mean{...}$ describes the mean value, and
$\bar{n}=N_{1}/A$ is the mean density of unemployed agents.  

\textbf{(iii)} For a derivation of Krugmans equation, we now have to
specify $f(r,t)$ in \eqn{stat}. Here, it is assumed that a single
unemployed agent which migrates due to \eqn{langev-red}, responds to the
\emph{total local income} in a specific way:
\begin{equation}
  \label{respo1}
  f(r,t)= \frac{\partial}{\partial r} \ln{[\omega(r)\,l(r,t)]}
\end{equation}
Eq. (\ref{respo1}) means that the migrant is guided by \emph{local
  gradients} in the \emph{total income}, with $\omega(r)$ being the local
income and $l(r,t)$ the local density of employed workers.
We note here again, that the migrant does not count on information about
the highest global income, he only ``knows'' about his vicinity. With
assumption \eqn{respo1}, we can rewrite \eqn{stat} using the discrete
notation, which is preferred by Krugman: 
\begin{equation}
  \label{stat-k}
  n^{stat}_{j}(t) =  \bar{n}\; \frac{\omega_{j}\;l_{j}(t)}{\sum_{j}
\omega_{j}\;l_{j}(t) \;/\;\sum_{j}}
\end{equation}
The corresponding equation (\ref{react2}) for $l_{j}(t)$ reads in the
discrete notation: 
\begin{equation}
  \label{react-disc}
  \frac{\partial}{\partial t}l_{j}(t)= k^{+}\,n_{j}(t)\;-\;k^{-}\,l_{j}(t)
\end{equation}
Since the Krugman equation deals with shares instead of densities, we have
to divide \eqn{react-disc} by $\sum_{j} l_{j}(t)$, which leads to:
 \begin{equation}
   \label{change-l}
   \frac{d}{d\,t}\lambda_{j}(t) = k^{-} \lambda_{j} \left(
\frac{k^{+}}{k^{-}} \frac{n_{j}}{l_{j}} -1 \right)
 \end{equation}
 
 \textbf{(iv)} Krugman assumes that the total number of employed workers
 is constant, so we use this assumption to replace the relation between
 the hiring and the firing rate in \eqn{change-l}. Eq. (\ref{react-disc})
 yields:
\begin{equation}
  \label{sum}
  \sum_{j} \frac{d}{d\,t}\lambda_{j}(t) = 0 \quad \Rightarrow \quad
\frac{k^{+}}{k^{-}}=\frac{\sum_{j} l_{j}}{\sum_{j} n_{j}}
\end{equation}
$n_{j}$ in \eqn{change-l} is now replaced by the quasistationary value,
\eqn{stat-k}.  Inserting further \eqn{sum} into \eqn{change-l} and using
the definitions of the average wage $\bar{\omega}$, \eqn{mean-w}, and
$\bar{n}=\sum_{j} n_{j}/\sum_{j}$, we finally arrive at:
\begin{equation}
  \label{krugman-2}
  \frac{d}{dt}\lambda_{j}(t) = \frac{k^{-}}{\bar{\omega}} \;
  \lambda_{j}\;(\omega_{j}-\bar{\omega})
\end{equation}
Eq. (\ref{krugman-2}) is identical with Krugmans law of motion of the
economy, \eqn{krugman}, if the prefactor $\rho$ in \eqn{krugman} is
identified as: $\rho=k^{-}/\bar{\omega}$. Hence, $\rho$ is a slowly
varying parameter, which depends both on the firing rate, which
determines the time scale, and on the average wage, $\bar{\omega}$, which
may change in the course of time.

It is an interesting question whether the assumptions (i)-(iv) which
lead to Krugmans law of motion in the economy, have some practical
evidence in an economic context. From a more theoretical perspective, it
is noteworthy that Krugmans equation, (\ref{krugman}), (\ref{krugman-2})
has an obvious analogy to a selection equation of the Eigen-Fisher type,
$\dot{x_{i}}=x_{i}(E_{i}-\mean{E_{i}})$. Here, $E_{i}$ is the fitness of
species $i$ and $\mean{E_{i}}$ is the mean fitness representing the
global selection pressure. It can be proved \cite{Feistel and Ebeling,
  1989} that this equation describes a competition process which finally
leads to a stable state with only \emph{one} surviving species. The
competition process may occur on a very long time scale, but
asymptotically a stable coexistence of many different species is
impossible.

In the economic context discussed by Krugman (1992), \eqn{krugman}
implies that finally \emph{all} workers are located in \emph{one} region
$j$, where the local real wage $\omega_{j}$ is equal to the mean real
wage, $\bar{\omega}$. However, in his paper, Krugman (1992) (see also
Krugman (1996b)) discusses computer simulations with 12 locations (on a
torus) which show the stable coexistence of two (sometimes three)
centers, roughly evenly spaced across the torus, with exactly the same
number of workers (cf. Fig. 7 in Krugman (1992) and Fig. 2.2. in Krugman
(1996b)). The paper does not provide information about the time scale of
the simulations, and stochastic influences (apart from a random initial
configuration) are not considered.

Krugmans model of spatial aggregation includes of course some more
complex economic assumptions, such as consideration of transportation
costs, distinction between agricultural and manifactured goods, price
index etc.  We are not going to discuss whether the stable coexistence of
two centers in Krugmans computer simulations might result from those
specific economic assumptions in his model or from the lack of
fluctuations (which could have revealed the instability of the
(deterministic) stationary state). Due to Krugman (1992), already a
slight variation in the parameters always led to a single center.

In the following, we will focus on the  more interesting question, 
whether the general dynamic model of migration and employment introduced
in Sect. 2, is able to produce a stable coexistence of different economic
centers under the presence of fluctuations. This may allow us to overcome
some of the shortages involved in Krugman equation. 

\section{Migration and Employment of Workers due to Wage Differences}
\subsection{Effective Diffusion}
The derivation of Krugmans equation was based on the assumtion (ii) that
the time scale for hiring and firing of workers is more determining than
the time scale of migration. If we explicitely consider unemployment in
our model, this assumption implies that there are always enough
unemployed workers which can be hired on demand. A growing economy,
however, might be determined just by the opposite limiting case: It is
important to attract workers/consumers to a certain area before the
output of production can be increased. Hence, the time scale for the
dynamics is determined by  migration processes. That means
the spatial distribution of the employed agents can be assumed in a
\emph{quasistationary equilibrium} compared to the spatial distribution
of the unemployed agents.
\begin{equation} 
\frac{\partial}{\partial t}l(r,t)= 0 \quad \Rightarrow \quad
l^{stat}(r,t)= \frac{k^{+}}{k^{-}}\;n(r,t)
\label{change-stat}
\end{equation}
Hence, the local density of employed agents can be expressed as a
function of the local density of the unemployed agents available, which
itself changes due to migration on a slower time scale.

We further assume that the unemployed agent who is able to migrate,
\eqn{langev-red}, responds to local gradients in the real wages (per
capita), instead of gradients in the total income, as assumed for
Krugmans equation.  The migrant tries to move
towards places with a higher wage, but again he only counts on
information in the vicinity. Hence, the guiding force $f(r_{i})$ is
determined as follows:
\begin{equation}
  \label{respo2}
  f(r,t)= \frac{\partial}{\partial r} \omega(r,t)
\end{equation}
For a further discussion, we need some assumptions about the local
distribution of the real wages, $\omega(r,t)$. It is reasonable to assume
that the local wages may be a functional of the local density of employed
agents, $\omega(r,t)=\omega\{l(r,t)\}$. Some specific assumptions about
this dependence will be discussed in the next section. With
\eqn{change-stat}, we can then rewrite the spatial derivative for the
wages as follows:
\begin{equation}
  \label{wage-dl}
  \frac{\partial \omega(r,t)}{\partial r}=\frac{\delta \omega\{l(r,t)\}}{%
\delta l}\;\frac{\partial l(r,t)}{\partial r}=
\frac{\delta \omega\{l(r,t)\}}{%
\delta l}\;\frac{k^{+}}{k^{-}}\;\frac{\partial n(r,t)}{\partial r}
\end{equation}
where $\delta$ denotes the functional derivative. 
Using the equations (\ref{change-stat}), (\ref{respo2}), (\ref{wage-dl}),
the Fokker-Planck equation for the change of the density of the
unemployed agents, \eqn{fpe-t} can now be rewritten as follows:
\begin{equation}
  \label{fpe-eff}
 \frac{\partial}{\partial t}\,n(r,t)\; = \;  
\frac{\partial}{\partial r} \left( D_{eff} 
\frac{\partial n(r,t)}{\partial r} \right) 
\end{equation}
The r.h.s. of \eqn{fpe-eff} now has the form of a usual
diffusion equation, with $D_{eff}$ being an \emph{effective diffusion
  coefficient}:
\begin{equation}
  \label{d-eff}
D_{eff}\;=\;D_{n} - \frac{k^{+}}{k^{-}}\;\frac{\delta \omega}{%
\delta l}\; n(r,t)
\end{equation}
Here, $D_{n}$ is the ``normal'' diffusion coefficient of the unemployed
agents, \eqn{fpe-t}. The additional terms reflect that the unbiased
diffusion is changed because of the response of the migrants to local
differences in the wage distribution. As we see from \eqn{d-eff}, there
are two contradicting forces determining the effective diffusion
coefficient: the normal diffusion, which keeps the unemployed agents
moving, and the response to the wage gradient. 

If the local wage decreases with the number of employed agents, then
\mbox{$\delta \omega/ \delta l < 0$}, and the effective diffusion increases.
That means unemployed agents migrate away from regions where employment
may result in an effective decrease of the marginal output. However, if
\mbox{$\delta \omega/ \delta l >0$}, then their wage effectively
\emph{increases} in regions with a larger number of employed agents, and
they are \emph{attracted} to these regions. As we see in \eqn{d-eff}, for
a certain \emph{positive feedback} between the local wage and the
employment density, the effective diffusion coefficient can be
\emph{locally negative} and unemployed agents do not leave these areas
once they are there. Economically speaking, these agents stay there,
because they may profit from the local increase in the employment
density.

We want to emphasize that this interesting dynamic behavior has been
derived \emph{without} any explicit economic assumptions. But for a
further discussion of the model, the three remaining functions which are
unspecified by now: (i) \mbox{$\delta \omega/ \delta l$}, (ii) $k^{+}$,
(iii) $k^{-}$, have to be specified, and that is of course where
economics comes into play.

\subsection{Determination of the Production Function}
In order to determine the economic functions, we refer to a
\emph{perfectly competitive industry} (where ``perfect'' means complete
or total). In this standard model \cite{see e.g. Case and Fair (1992)},
the economic system is composed of many firms, each small relative to the
size of the industry. These firms represent seperate economies sharing
common pools of labor and capital.  New competitors can freely enter/exit
the market, hence the number of production centers is not limited or
fixed.  Further, it is assumed that every firm produces the same (one)
product and every firm uses only \emph{one} variable input. Then, the
\emph{maximum profit condition} tells us that ``firms will add inputs as
long as the marginal revenue product of that input exceeds its market
price.''  In the case of labor as variable input, the price of labor is
the wage, and a profit maximizing firm will hire workers as long a the
marginal revenue product exceeds the wage of the added worker.

The marginal revenue product $MRP= MP\times P$ is the additional revenue
a firm earns by employing one additional unit of input, where $P$ is the
price of output and $MP$ is the marginal product. In a perfectly
competitive industry, however, no single firm has any control over
prices. Instead, the price results from the interaction of many suppliers
and many demanders. In a perfectly competitive industry, every firm sells
its output at the market equilibrium price, which is simply normalized to
one, hence the marginal revenue product is determined by $MP$.  

The marginal product, $MP$, can be derived from a \emph{production
  func\-tion} - \mbox{$Y(r,t) = A(r,t)\,g(r,t)$}, which describes the
relationship between inputs and outputs (e.i. the technology of
production). Usually, the production function may include the effect of
different inputs, such as capital, public goods or natural resources.
Here, we concentrate only on one variable input, labor. Thus, $g(r,t)$ is
assumed a Cobb-Douglas production function $l^{\beta}(r,t)$ with a common
(across regions) exponent $\beta$, $l(r,t)$ being the local density of
employees. The exponent $\beta$ describes how a firms output depends on
the scale of operations. Increasing returns to scale are characterized by
a lower average cost with an increasing scale of production, hence
$\beta>1$ in this case. On the other hand, if an increasing scale of
production leads to higher average costs (or lower average output), we
have the situation of decreasing returns to scale, with $\beta<1$. In the
following, we will restrict the discussion to $\beta<1$, common to all
regions.

The prefactor $A(r,t)$ represents economic details of the level of
productivity. We assume that these influences can be described by two
terms: \mbox{$A(r,t)=A_{c}+A_{u}$}. %
$A_{c}=\mbox{const.}$ should summarize those output dependences on
capital, resources etc., which are not explicitely discussed here. The
new second term, $A_{u}$, considers \emph{cooperative effects} which
result from interactions among the workers. Since all cooperative effects
are \emph{non-linear} effects, $A_{u}$ should be a non-linear functional
of $l(r,t)$: $A_{u}=A_{u}\left\{l(r,t)\right\}$.

Hence, the production function depends explicitely only on $l(r,t)$, now:
\begin{equation}
  \label{prod1}
 Y\{l(r,t)\} = \Big[A_{c} +A_{u}\{l(r,t)\}\Big] \;l^{\beta}(r,t) 
\end{equation}
The marginal product ($MP$) is the additional output by adding one more
unit of input, i.e.  $MP=dY/dl$, if labor is input. The wage of a
potential worker will be the marginal product of labor:
\begin{equation}
  \label{wage}
  w\{l(r,t)\}= \frac{\delta Y\{l(r,t)\}}{\delta l}
\end{equation}
If a firm faces a market wage rate of $\omega^{\star}$ (which could be
bound e.g. by minimum wage laws), then, in accordance with the maximum
profit condition, a firm will hire workers as long as:
\begin{equation}
  \label{wage-star}
  \frac{\delta Y\{l(r,t)\}}{\delta l} >  \omega^{\star}
\end{equation}
To complete our setup, we need to discuss how the prefactor $A_{u}$
depends on the density of employees, $l(r,t)$. Here, we assume that
the cooperative effects will have an effect only in the intermediate
range of $l$. For small production centers, the synergetic effect
resulting from the mutual stimulation among the workers is too low. On
the other hand, for very large production centers, the advantages of the
cooperative effects might be compensated by the disadvantages of the
massing of agents.
Thus we will assure, that $Y \sim l^{\beta}$ both in
the limit $l \to 0$ and $l \to \infty$.  These
assumtions are concluded in the ansatz:
\begin{equation}
  \label{a-u}
A_{u}\left\{l(r,t)\right\} \sim \exp\Big\{u\left\{l(r,t)\right\}\Big\}
\end{equation}
where the utility function $u(l)$ describes the mutual stimulation among
the workers in powers of $l$:
\begin{equation}
  \label{u-l}
u(l)=a_{0}+a_{1}l+a_{2}l^{2}+...  
\end{equation}
The series will be truncated after the second order. The constants
$a_{i}$ characterize the effect of cooperation, with $a_{0}>0$.
Especially, the case $a_{1}>0,a_{2}<0$ considers saturation effects in
the cooperation, e.i. the advantages of cooperation will be
compensated by disadvantages of crowding. This idea implies that there is
an optimal size for taking advantages of the cooperative effect, which is
determined by the ratio of $a_{1}$ and $a_{2}$. If one believes that
cooperative effects are an always increasing function in $l$, then simply
$a_{2}>0$ can be assumed.

If cooperative effects are neglected, i.e. $a_{1}=a_{2}=0$, then we
should obtain the ``normal'' production function,
$Y(l)=\bar{A}l^{\beta}$. The constant $\bar{A}$ is determined by a
relation between $A_{c}$ and $a_{0}$ which results from the
eqs. (\ref{prod1}), 
(\ref{a-u}). Within our approach, the
constant $A_{c}$ is not specified. If we, without
restrictions of the general case, choose:
\begin{equation}
  \label{bar-a}
 \bar{A}=A_{c}+\exp\{a_{0}\} \equiv  \;2 A_{c}  
\end{equation}
then the production function, $Y(l)$, \eqn{prod1}, with respect to
cooperative effects can be expressed as follows:
\begin{equation}
  \label{prod2}
 Y\left\{l(r,t)\right\}= \frac{ \bar{A}}{2}
\left[1+ \exp\left(a_{1}\,l+a_{2}\,l^{2}\right) \right]\;l^{\beta}
\end{equation}
Fig. 1 presents the production function, \eqn{prod2} dependent on the
density of employees, which can be compared with the ``normal''
production function, $Y(l)=\bar{A}l^{\beta}$. 
\begin{figure}[htbp]
\label{fig1}
\centerline{\psfig{figure=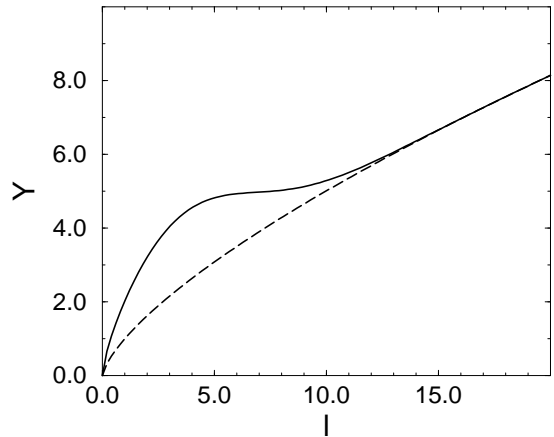,width=8cm}}
\caption[fig1]{
  Production function $Y(l)$ vs. density of employed workers, $l$. The
  solid line results from \eqn{prod2} (parameters: $\bar{A}=2$,
  $a_{1}=0.06$, $a_{2}=-0.035$, $\beta=0.7$). The dashed line shows the
  production function \emph{without} cooperative effects
  ($a_{1}=a_{2}=0$).}
\end{figure}
Clearly, we see an increase in the total output due to cooperation
effects among the workers. If we assume $a_{1}>0, a_{2}<0$, this increase
has a remarkable effect only in an intermediate range in $l$.  Both for
$l \to 0$ and $l \to \infty$ the cooperative effect vanishes.

Once the production function is determined, we also have determined the
local wage $\omega\{l(r,t)\}$, \eqn{wage}, as a function of the
density of employees: 
\begin{eqnarray}
  \label{wage2}
  \omega\left\{l(r,t)\right\} &=&
 \frac{ \bar{A}}{2} \left[1+ \exp\left(a_{1}\,l+a_{2}\,l^{2}\right) 
\right]\;\beta \,l^{\beta-1} \nonumber \\
&+& \frac{ \bar{A}}{2} \;\exp\left(a_{1}\,l+a_{2}\,l^{2}\right) 
\,(a_{1}+2a_{2}l)\;l^{\beta} 
\end{eqnarray}
Hence, also the derivative
$\delta\omega/\delta l$, used for the effective
diffusion coefficient, $D_{eff}$, \eqn{d-eff} is determined. Fig. 2 shows
both functions dependent on the density of employees.
\begin{figure}[htbp]
\label{fig2}
\centerline{%
\psfig{figure=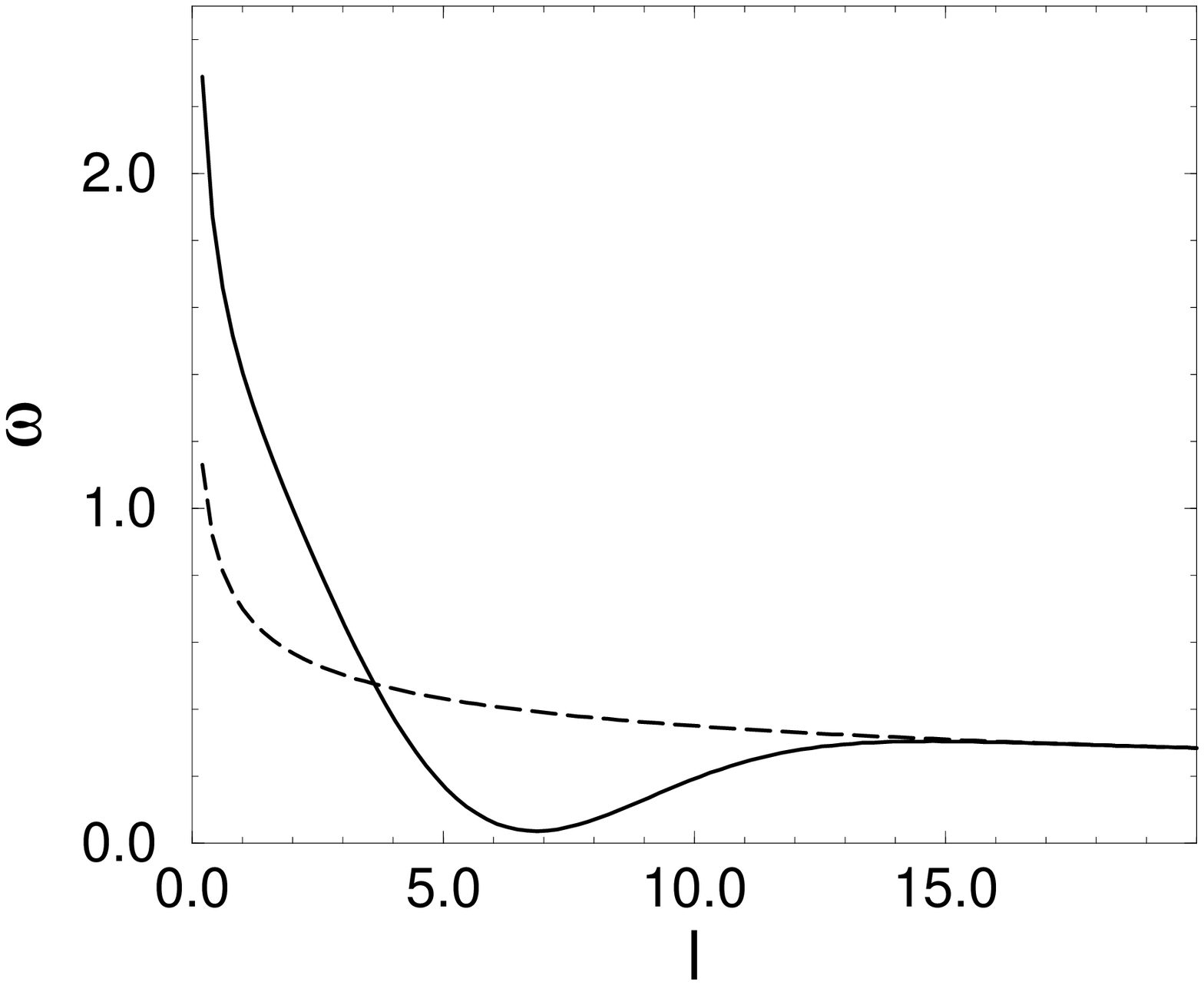,width=8cm}\hfill
\psfig{figure=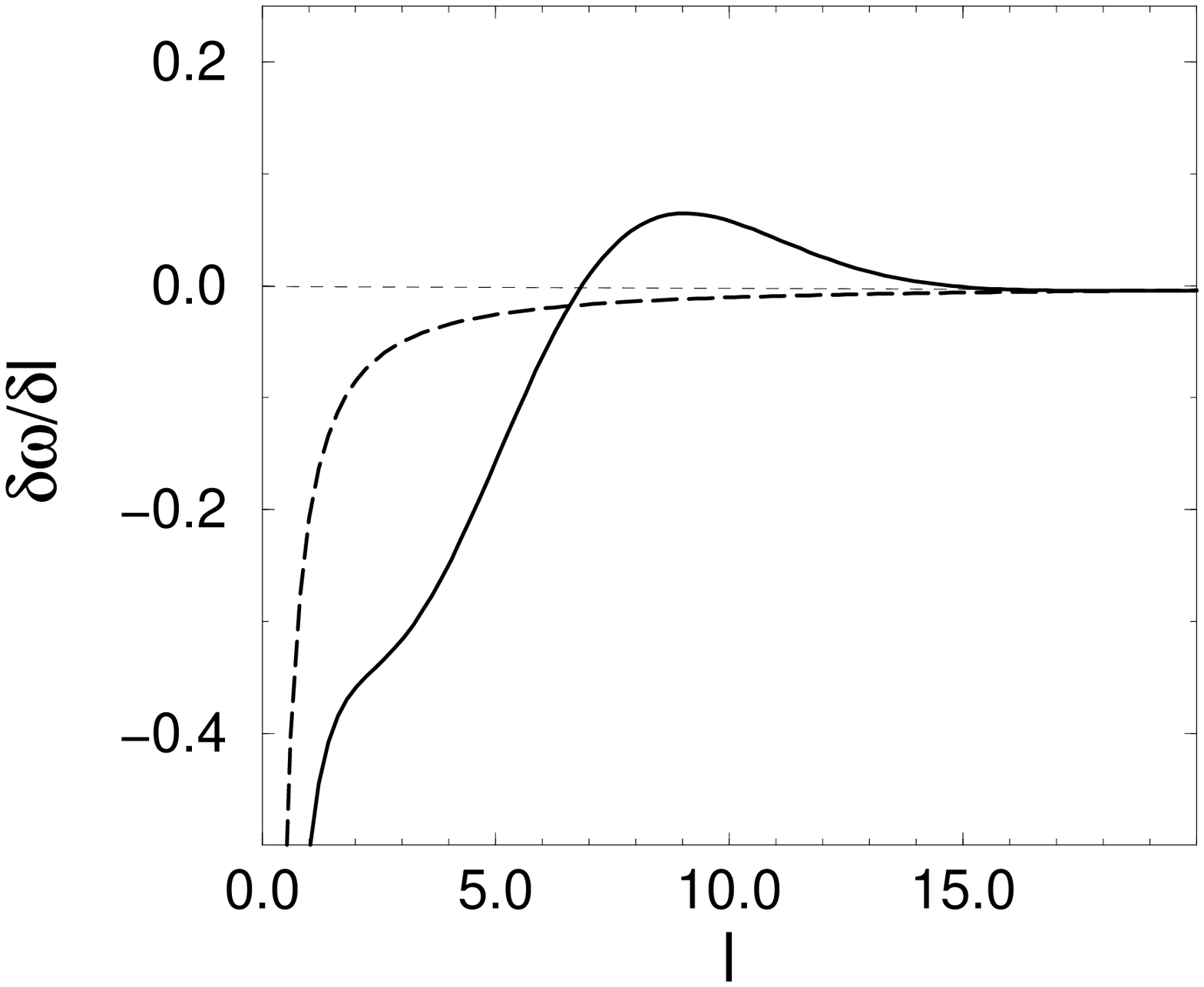,width=8cm}}
\caption[fig2]{%
  (left) Wage $\omega$, \eqn{wage2}, (right) derivative
  $\delta \omega/\delta l$ vs. density of employed workers, $l$. The
  solid lines show the functions \emph{with}, the dashed lines
  \emph{without} cooperative effects (for the parameters see Fig. 1).}
\end{figure}

The left part of Fig. 2 indicates that, within  a certain range of $l$,
the derivative $\delta\omega/\delta l$ could be indeed positive, due
to cooperative effects. With respect to the discussion in Sect. 4.1.,
this means that the effective diffusion coefficient, \eqn{d-eff}, can be
possibly negative, i.e. unemployed workers will stay in these regions. 

\subsection{Determination of the Transition Rates}
The transition rate $k^{+}$ for hiring unemployed workers is implecitely
already given by the conclusion, that firms hire workers as long as the
marginal revenue product exceeds the wage of the worker. In accordance
with \eqn{wage-star} we define: 
\begin{equation}
  \label{kplus}
  k^{+} = k^{+}\{l(r,t)\} 
=\alpha \; \exp{\left\{ \frac{\D \delta
  Y\left\{l(r,t)\right\}}{\D \delta l} - \omega^{\star} \right\} }
\end{equation}
Here, $\alpha$ determines the time scale of the transitions. $k^{+}$,
which is now a function of the local economic situation, is significant
larger than the level of random transitions, represented by $\alpha$,
only if the maximum profit condition allows hiring. Otherwise, the
hiring rate tends to zero. 

The firing rate $k^{-}$ could be simply determined opposite to
$k^{+}$. Then, from the perspective of the employee, firing is caused by
the local situation of the economy, i.e. by \emph{external reasons}:
workers are fired if $\delta Y/\delta l < \omega^{\star}$.  A more refined
description, however, should consider also \emph{internal reasons}: a
worker cannot only \emph{loose} his job, he can also \emph{quit} his job
himself for better opportunities, e.g. because he wants to move to a
place where he earns a higher wage. 

It is reasonable to assume that the internal reasons depend again on
spatial gradients in the wage. Due to \eqn{langev-red} the unemployed
agent migrates while guided by gradients in the wage. The employed agent
at the same location may have the same information. Noteworthy again,
this is only a local information about differences in the wage. If the
local gradients in the wage are small, the internal reasons to quit the
job vanish, and the firing depends entirely on the (external) economic
situation. However, if these differences are large, the employee might
quit his job for better chances. 

We note that the latter process was already considered in Krugmans law of
motion in the economy, Sect. 3. Different from the assumptions involved
in Krugmans \eqn{krugman}, here the process: employment $\to$
unemployment $\to$ migration $\to$ reemployment is \emph{explicitely}
modeled. It does not occur with infinite velocity, and there is no
guarantee for reemployment. 

Hence, we define the ``firing rate'' which describes the transition from
an employed to an unemployed agent, as follows: 
\begin{equation}
  \label{kminus}
  k^{-} =  k^{-}\{l(r,t)\} =\alpha \; \exp{\left\{ -\left[
\frac{\D \delta
      Y\left\{l(r,t)\right\}}{\D \delta l} - \omega^{\star} \right] 
+\; q\, \frac{\D \partial \omega(r)}{\D \partial r} \right\}}
\end{equation}
The additional parameter $q$ can be used to weight the influence of
spatial gradients on the employee. 

Eventually, in this section we have determined the variables $f(r,t)$,
$k^{+}$ and $k^{-}$ via a production function $Y\{l(r,t)\}$, which
represents certain economic assumptions. Now, we can turn back to the
dynamic model described in Sect. 2, which now can be solved by means of
computer simulations.

\section{Numerical Simulations}
\subsection{Stochastic Simulation Technique}
Before presenting the results of the computer simulations, the simulation
technique should be shortly discussed. The computer program has to deal
with three different processes, which have to be discretized in time for
the simulation: (i) the movement of active Brownian particles with
$\theta_{i}=1$ due to the overdamped Langevin \eqn{langev-red}, (ii) the
transition of the particles due to the rates, $k^{+}$, \eqn{kplus}, $
k^{-}$, \eqn{kminus}, and (iii) the generation of the field
$\omega(r,t)$.

Considering \eqn{langev-red}, the new $x$-position of a particle $i$ with
$\theta_{i}=1$ on the two dimensional surface at time $t+\Delta t$ is
given by:
\begin{equation}
\label{disc-x}
 x_{i}(t+\Delta t) = x_{i}(t) + 
\left.\frac{\partial \omega}{\partial x} \right|_{x_{i}}\;\Delta t+
  \sqrt{2D_n \Delta t}\; \mbox{GRND}
\end{equation}
The equation for the $y$-position reads accordingly. $\Delta t$ is the
\emph{non-constant} time step, which is calculated below. $D_{n}=\eps$ is
the diffusion coefficient, and \emph{ GRND} is a Gaussian random number
with mean equals zero and standard deviation equals unity.

In oder to calculate the spatial gradient of the wage field $\omega$,
\eqn{wage2}, we
have to consider its dependence on $l(r,t)$,  which is a local
density. The density of employed agents is calculated assuming that the
surface is divided into boxes with the spatial (discrete) indices $u,v$
and unit length $\Delta s$. Then, the local density is given by the
number of agents with $\theta_{i}=0$ inside a box of size $(\Delta
s)^{2}$:
\begin{equation}
  \label{disc-l}
  l(r,t) \; \Rightarrow \; l_{uv}(t) = \frac{1}{(\Delta s)^{2}} 
\int_{x_{u}}^{x_{u}+\Delta s} \int_{y_{v}}^{y_{v}+\Delta s} 
dx'\, dy'\; C_{0}(x',y') 
\end{equation}
We note that $\Delta s \gg \Delta x$, with $\Delta x = x(t+\Delta t)
-x(t)$ being the spatial move of the migrating agent into $x$-direction
during time step $\Delta t$. That means that the migration process is
really simulated as a motion of the agents on a two dimensional plane,
rather than a hopping process between boxes.

Using the box coordinates $u,v$, the production function and the wage
field can be rewritten as follows:
\begin{equation}
  \label{disc-Y-w}
  Y\left\{l(r,t)\right\} \;\Rightarrow \; 
  Y\left\{l_{uv}(t)\right\} = Y_{uv}(t) \;;\;\;
\omega(r) \; \Rightarrow \;
\omega_{uv}=\frac{\delta Y_{uv}(t)}{\delta l}
\end{equation}
The spatial gradient is then defined as: 
\begin{equation}
  \label{disc-grad}
  \frac {\partial \omega_{uv}}{\partial x} = 
\frac{\omega_{u-1,v} - \omega_{u+1,v}}{2\Delta s}  \;;\;\;
 \frac {\partial \omega_{uv}}{\partial y} = 
\frac{\omega_{u,v-1} - \omega_{u,v+1}}{2\Delta s}
\end{equation}
where the indices $u-1$, $u+1$, $v-1$, $v+1$ refer to the left, right,
lower and upper boxes adjacent to box $uv$. We further note that for the
simulations \emph{periodic boundary conditions} are used, therefore the
neighboring box is always specified. 

Using the discretized versions, \eqn{disc-l}, (\ref{disc-Y-w}),
(\ref{disc-grad}), the transition rates can be reformulated as
$k_{uv}^{+}$, $k_{uv}^{-}$ accordingly. They determine the \emph{average
  number} of transitions of a particle in the internal state
$\theta_{i}$, located in box $uv$, during the next time step, $\Delta t$.
In a stochastic simulation however, the actual number of reactions is a
\emph{stochastic} variable, and we have to assure that the stochastic
number of reactions (i) does not exceed the actual number of particles
available during $\Delta t$, and (ii) is equal to the average number of
reactions in the limit $t \to \infty$.

This problem can be solved by using the stochastic simulation technique
for reactions, which defines the appropriate time step, $\Delta t$, as a
random variable. Let us assume that we have exactly $N_{1}$ particles
with the internal state $\theta_{i}=1$ in the system at time $t=t_{0}$. 
Then the probability $P(N_{1},t_{0})$ equals one, and the probability for
any other number $N_{1}'$ is zero. With this intial condition, the
master equation to change $N_{1}$ reads:
\begin{equation}
  \label{master-1}
  \frac{\partial P(N_{1},t)}{\partial t}= \sum_{i=1}^{N}
  (k_{i}^{+}+k_{i}^{-}) \,P(N_{1},t)\;\;;\quad P(N_{1},t_{0})=1
\end{equation}
where $N$ is the total number of particles, and $k_{i}^{+}$, $k_{i}^{-}$
are the transition rates for each particle. The solution of this equation
yields:
\begin{eqnarray}
  \label{m-solv}
 P(N_{1},t-t_{0})& \sim& \exp\left(-\frac{t-t_{0}}{t_{m}}\right) 
\nonumber \\
t_{m} &=&\frac{1}{\sum\limits_{i=1}^{N} k_{i}} \;\;\;; \quad
\sum\limits_{i=1}^{N} k_{i} = \sum\limits_{i=1}^{N_{1}} k_{i}^{+} +
\sum\limits_{i=N_{1}+1}^{N} k_{i}^{-}
\end{eqnarray}
Here, $t_{m}$ is the mean life time of the state $N_{1}$. For $t-t_{0} \ll
t_{m}$, the probability $P(N_{1},t)$ to find still $N_{1}$ is almost one,
but for $t-t_{0}\gg t_{m}$ this probability goes to zero. The time when
the change of $N_{1}$ occurs, is most likely about the mean life time,
$t-t_{0}\approx t_{m}$. In a stochastic process, however, this time
varies, hence, the \emph{real life time} $t-t_{0}=\tau$ is a randomly
distributed variable. Since we know that $P(N_{1},t)$ has values between
$[0,1]$, we find from eq. (\ref{m-solv}):
\begin{equation}
  \label{tau}
  \tau = t-t_{0}= -\,t_{m}\, \ln\Big\{RND[0,1]\Big\}
\end{equation}
(Zero actually has to be excluded.). That means, after the real life time
$\tau$, one of the possible processes which change $N_{1}$ occurs. Each
of these processes has the probability:
\begin{eqnarray}
  \label{wk}
p(\theta_{i}=0,t\to\theta_{i}=1,t+\tau) &=&
k_{i}^{-} / \sum\limits_{i=1}^{N} k_{i}\nonumber\\
p(\theta_{i}=1,t\to\theta_{i}=0,t+\tau)&=&
k_{i}^{+} / \sum\limits_{i=1}^{N} k_{i} 
\end{eqnarray}
Thus, with a second random number \emph{RND}[0,$\sum k_{i}$] it will be
determined which of these possible processes occurs. It will be the
process $q$ which satisfies the condition:
\begin{equation}
  \label{prob-r}
\sum\limits_{i=1}^{q-1} k_{i} < RND \;\;;\quad
\sum\limits_{i=1}^{q} k_{i} >  RND
\end{equation}
For the transition probabilities to change $N_{1}$, we find in
particular:
\begin{eqnarray}
  \label{prob}
Prob[N_{1},t\to N_{1}+1,t+\tau] & = & \sum_{i=1}^{N_{0}}  k^{-}_{i} 
t_{m} \nonumber \\ 
Prob[N_{1},t\to N_{1}-1,t+\tau] & = & \sum_{i=1}^{N_{1}}  
k^{+}_{i} t_{m}
\end{eqnarray}
Obviously, the sum over these probabilities is one, which means, during
the time intervall $\tau$ one of these processes occurs with certainity.
Further, using the definition of $\tau$, we see that 
\begin{equation}
  \label{comp}
t =  \sum_{l=1}^{z}t_{m}^{l}= \sum_{l=1}^{z}\tau^{l}
\end{equation}
yields asymptotically, i.e. after a large number of simulation
steps, $z \gg1$.

Hence, determining the time step as $\Delta t=\tau$, \eqn{tau}, ensures
that only one transition occurs during one time step and the number of
transitions does not get out of control.  Further, in the asympotitc
limit, the actual (stochastic) number of transitions is equal the average
number.  The (numerical) disadvantage might be that the time step $\Delta
t$ is not constant, so it has to be recalculated after each cycle before
moving the particles with respect to \eqn{disc-x}. This may slow down the
speed of the simulations considerably.

Let us conclude the procedure to simulate the movement and the transition
of the particles:
\begin{enumerate}
\item calculate the density of particles $l_{uv}(t)$, \eqn{disc-l}
\item calculate the production function $Y_{uv}(t)$ and the wage field
  $w_{uv}(t)$, \eqn{disc-Y-w}
\item calculate the sum over all possible transitions, 
  which determines the mean life time $t_{m}$, \eqn{m-solv}, of the
  current system state
\item calculate from the mean life time the \emph{actual} time step
  $\Delta t=\tau$, \eqn{tau} by drawing a random number
\item move all particles with $\theta_{i}=1$ according to the Langevin
  equation (\ref{disc-x}) using the time step $\tau$
\item calculate which one of the particles undergoes a transition by
  drawing a second random number, eq. (\ref{prob-r})
\item update the system time: $t=t+\tau$ and continue with 1.
\end{enumerate}

\subsection{Computer Simulations of Spatial Economic Aggregation}
With the determination of the three functions $f(r,t)$, $k^{+}$ and
$k^{-}$ with respect to some economic assumptions, we have completed
our dynamic model, described in Sects. 2 and 4. 
In this section, we want to discuss some features of the dynamics by
means of a computer with $500$ active Brownian particles. Initially,
every particle is randomly assigned an internal parameter (either $0$ or
$1$), and a position on the surface, which has been divided into $10
\times 10$ boxes of unit length $\Delta s=1$.
The diffusion coefficient, which describes the mobility of the migrants,
is set to $0.01$ (in arbitrary units), So, a simple Brownian particle
would approximately need a time of $t=50$ for a mean spatial displacement of $1$
(which is the spatial extension of a box).
The minimum wage $\omega^{\star}$ is set to $0.1$. Further,
$\alpha=1$ and $q=1$, for the remaining parameters see Fig. 1.

In the following we will restrict the discussion to the spatio-temporal
evolution of the densities of employed and unemployed agents. Other
quantities of interest, such as the spatio-temporal wage distribution,
the production function and the local values of the ``hiring'' and
``firing'' rates have been of course calculated for the simulation, but
will be discussed in a subsequent paper, which also investigates
different parameter sets.

Fig. 3 shows snapshots of the evolution of the spatial density of
employed agents, while Fig. 4 shows the corresponding spatial density of
unemployed agents. Fig. 5 presents the evolution of the total number of
employed and unemployed agents, in terms of the total share
$x_{\theta}=N_{\theta}/N$ with respect to \eqn{total}.
\begin{figure}[htbp]
\label{fig3}
\centerline{
\psfig{figure=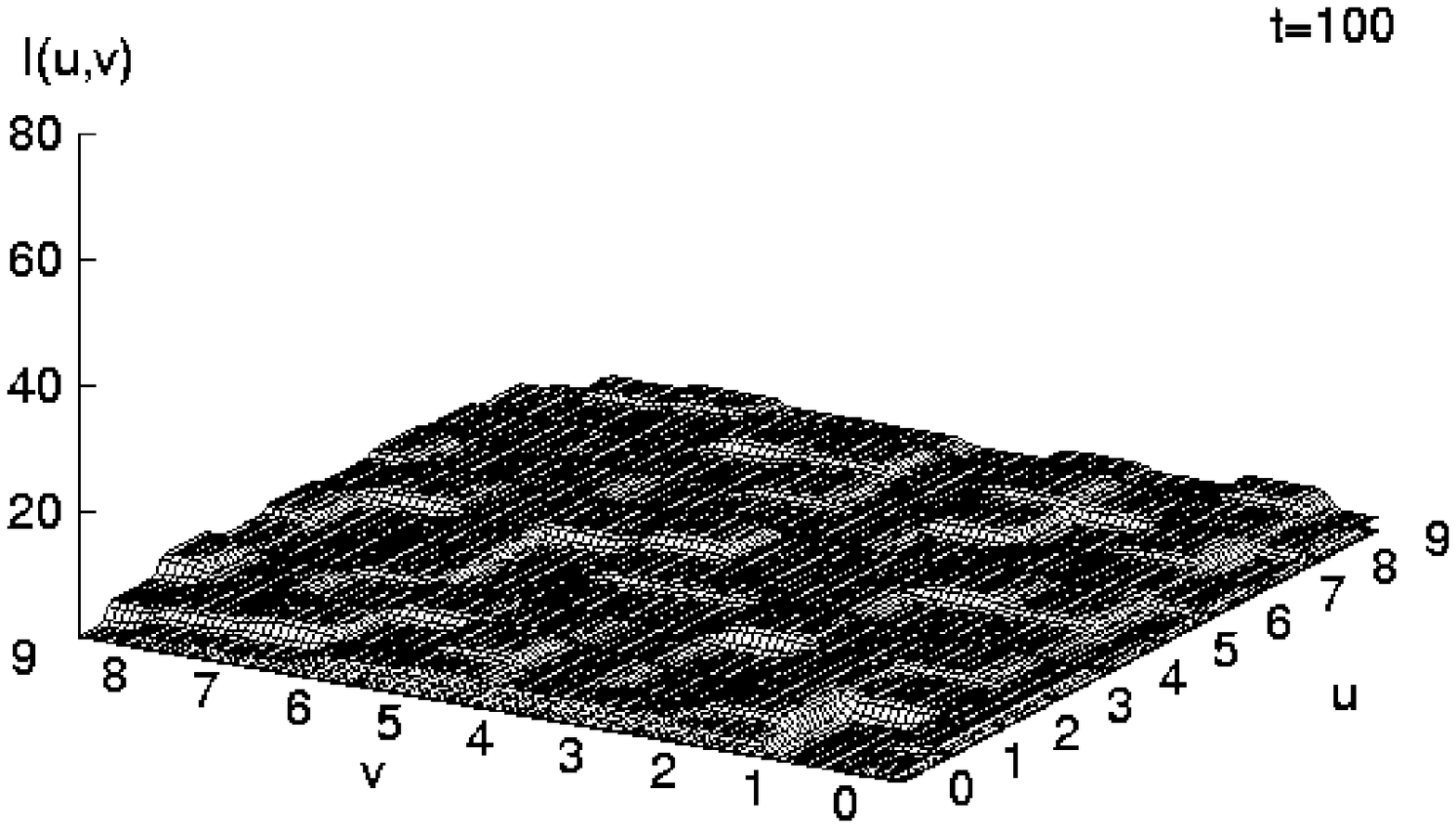,width=8.8cm,angle=-90}\hfill
\psfig{figure=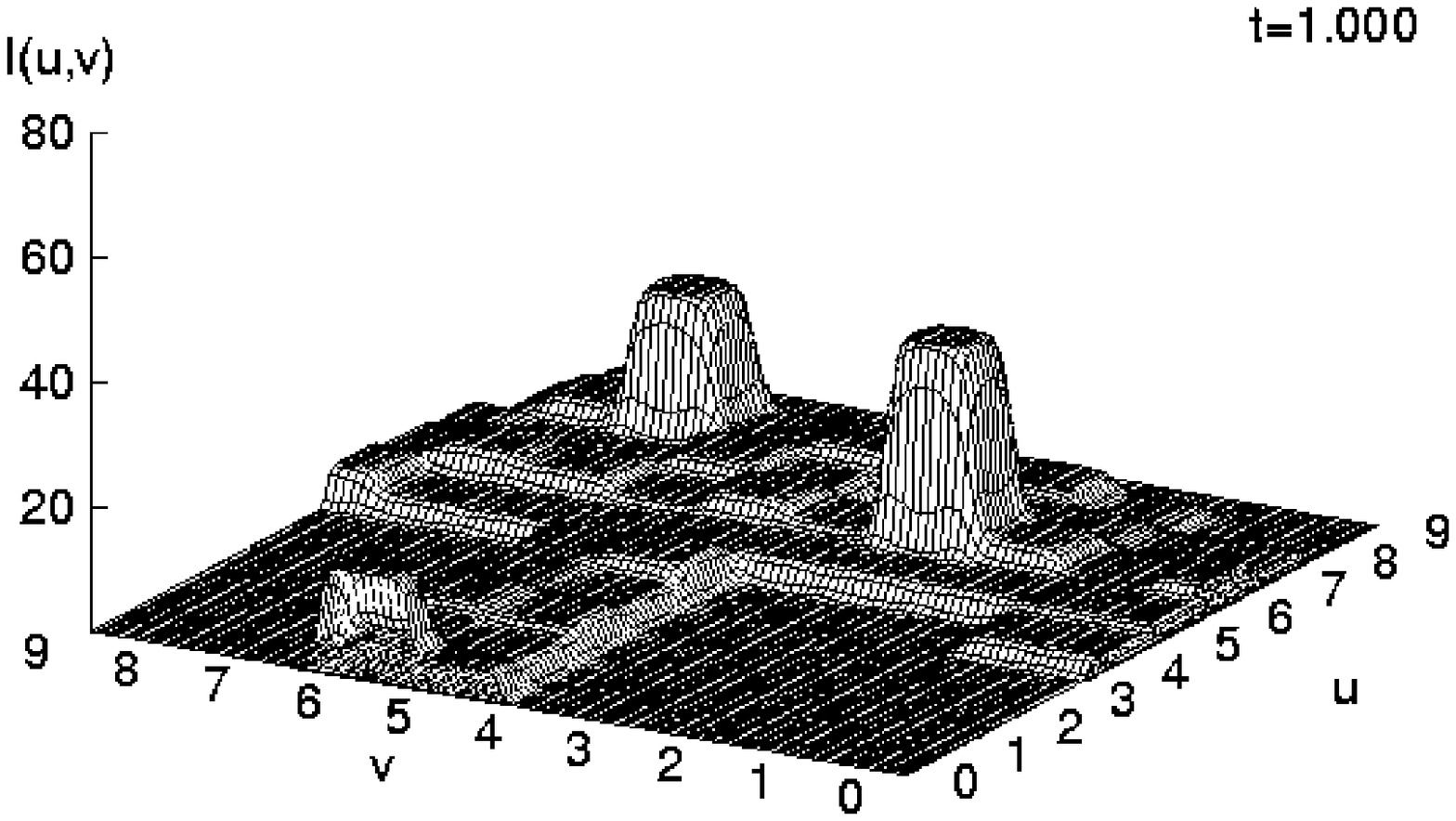,width=8.8cm,angle=-90}}
\centerline{
\psfig{figure=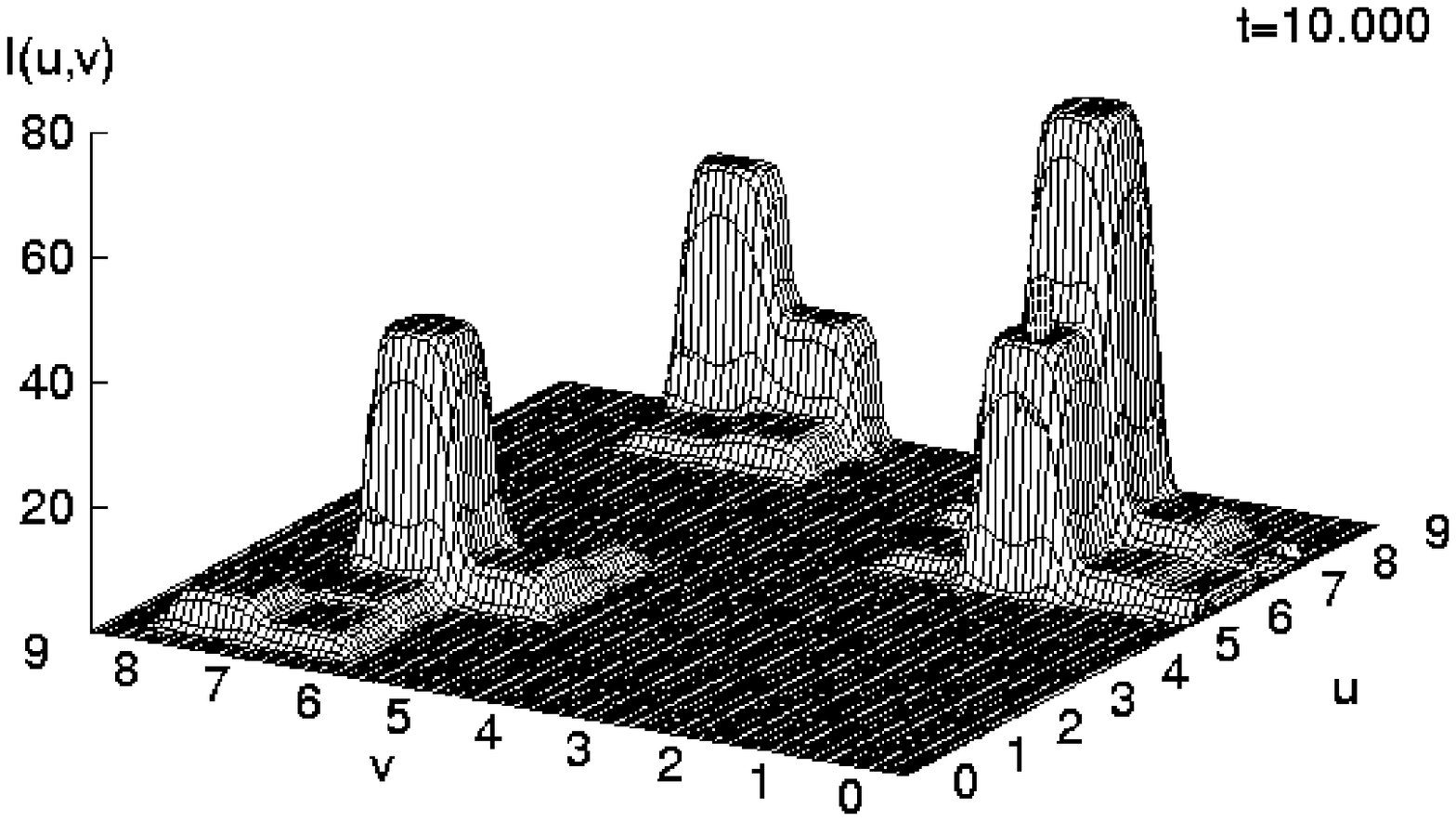,width=8.8cm,angle=-90}\hfill
\psfig{figure=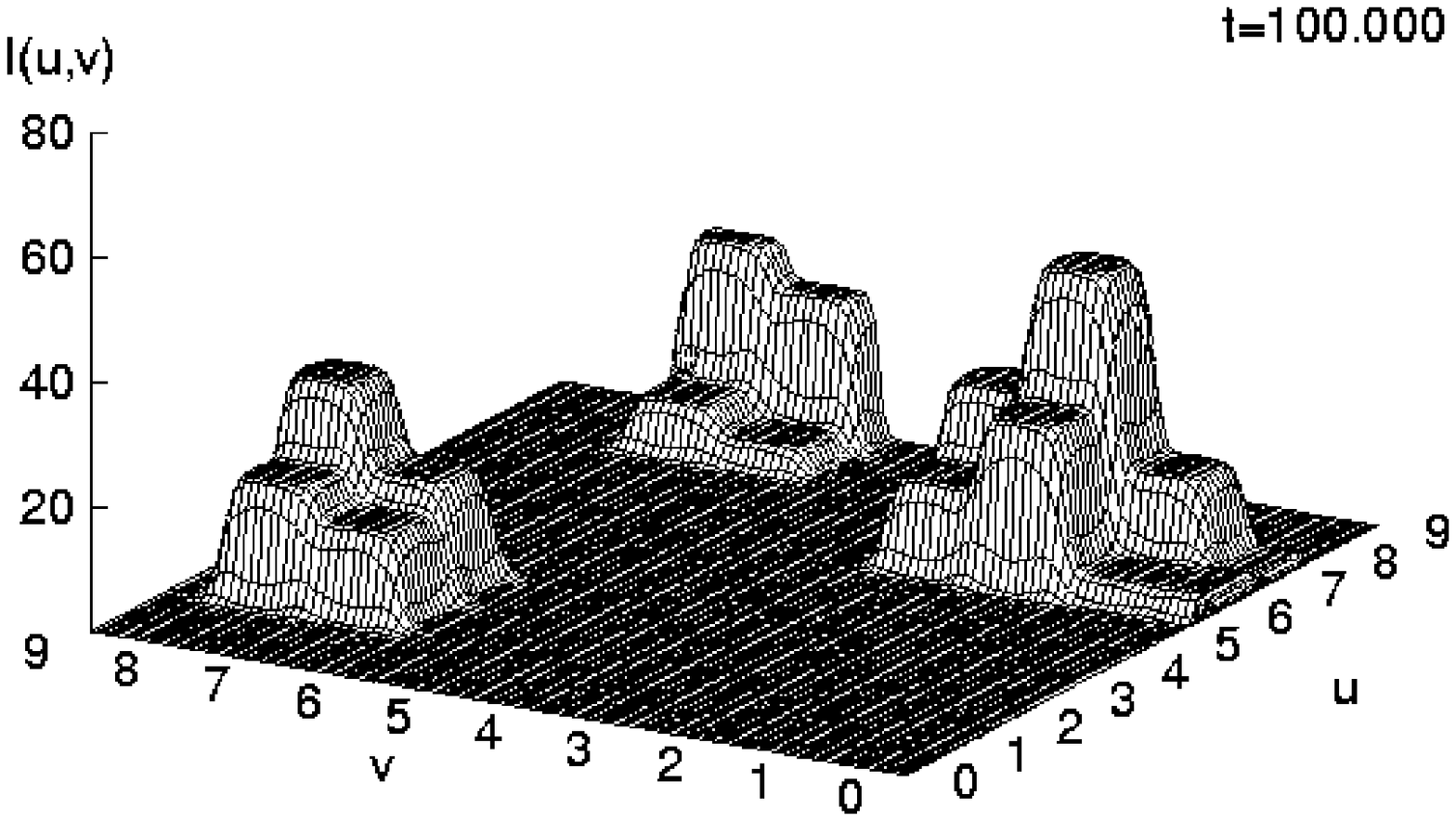,width=8.8cm,angle=-90}}
\caption[fig3]{%
Spatial density of employed agents, $l(u,v)$, for different times $t$
(for the parameters and initial conditions of the simulation, see text).}
\end{figure}

\begin{figure}[htbp]
\label{fig4}
\centerline{
\psfig{figure=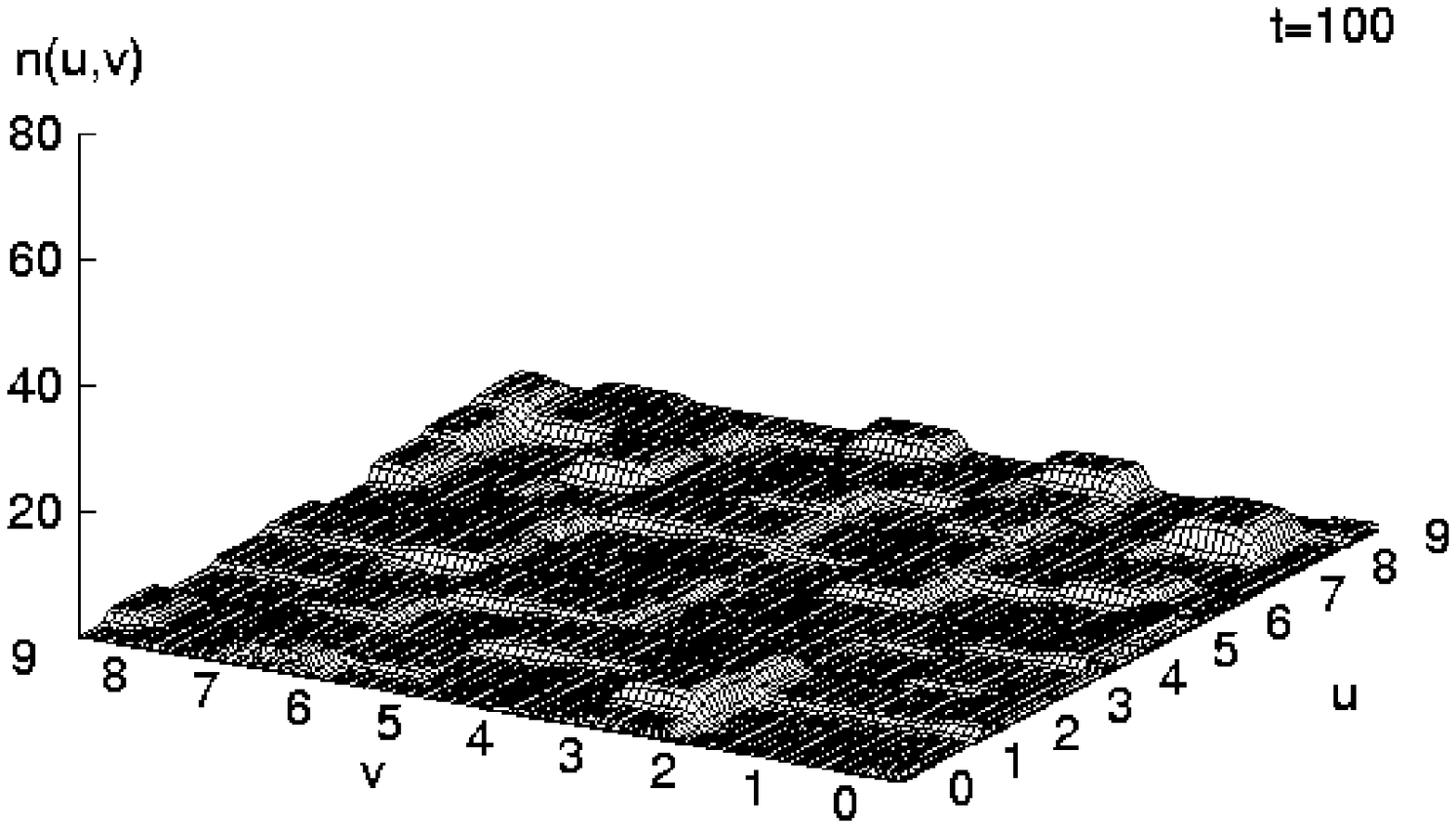,width=8.8cm,angle=-90}\hfill
\psfig{figure=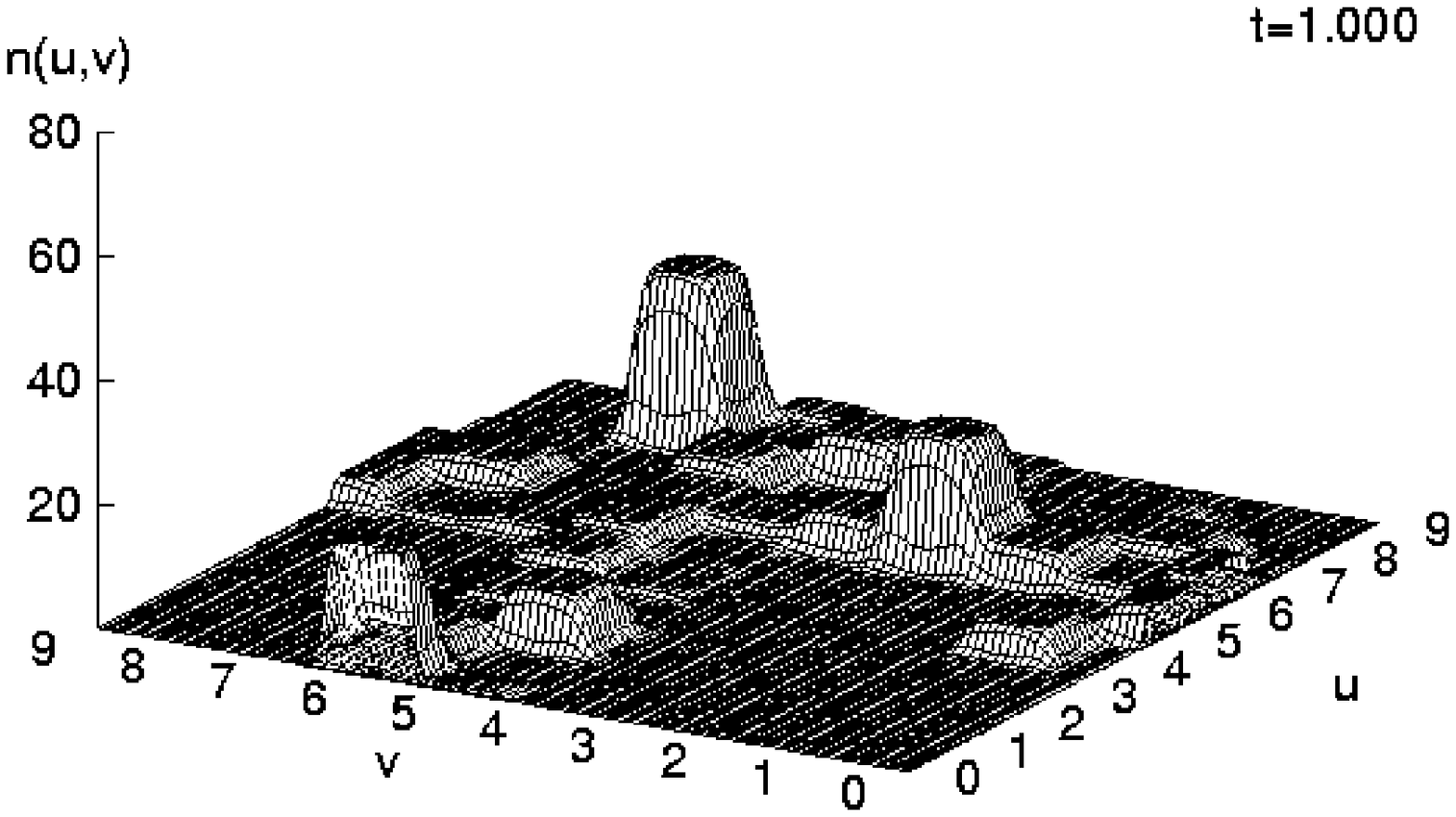,width=8.8cm,angle=-90}}
\centerline{
\psfig{figure=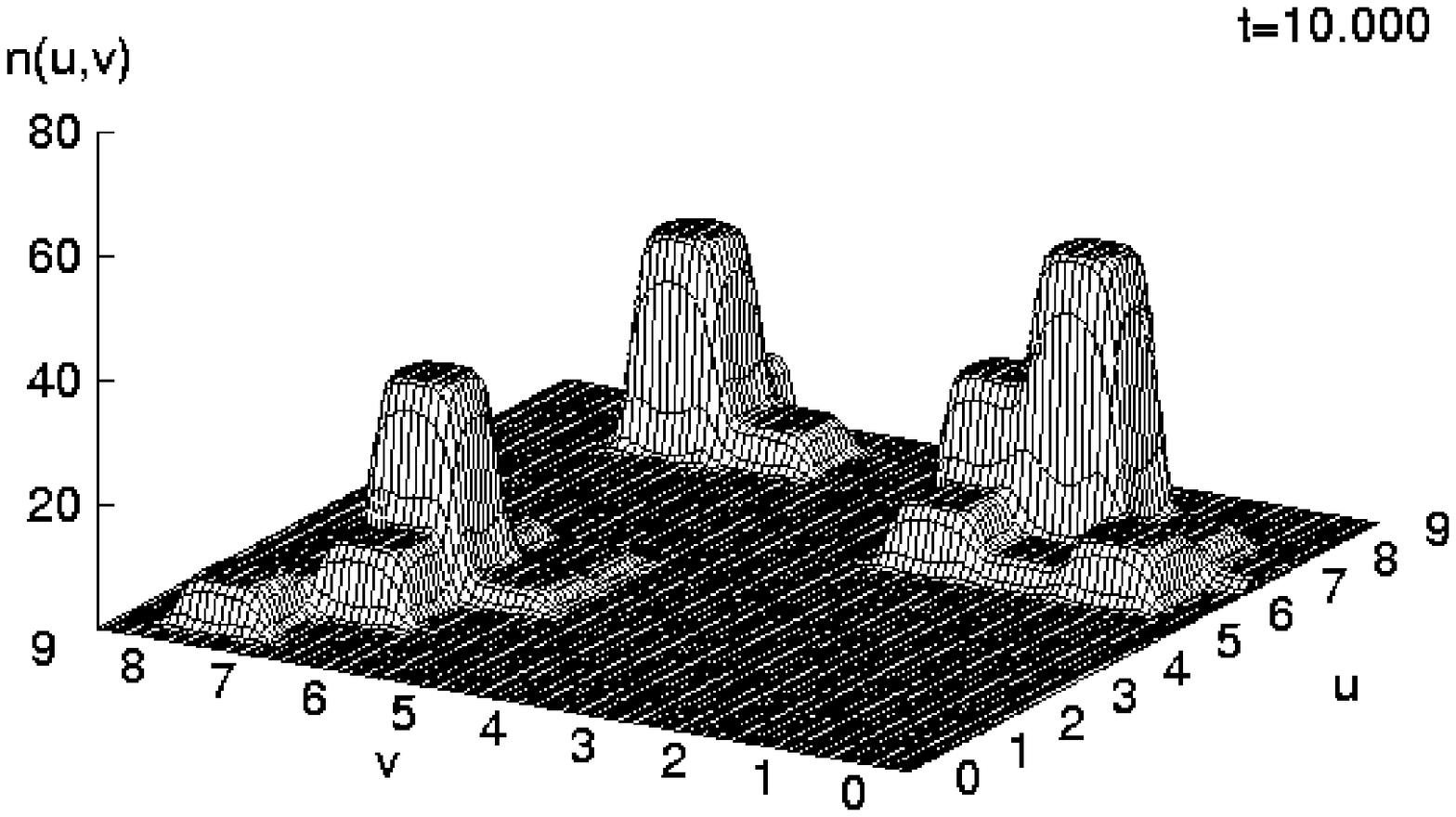,width=8.8cm,angle=-90}\hfill
\psfig{figure=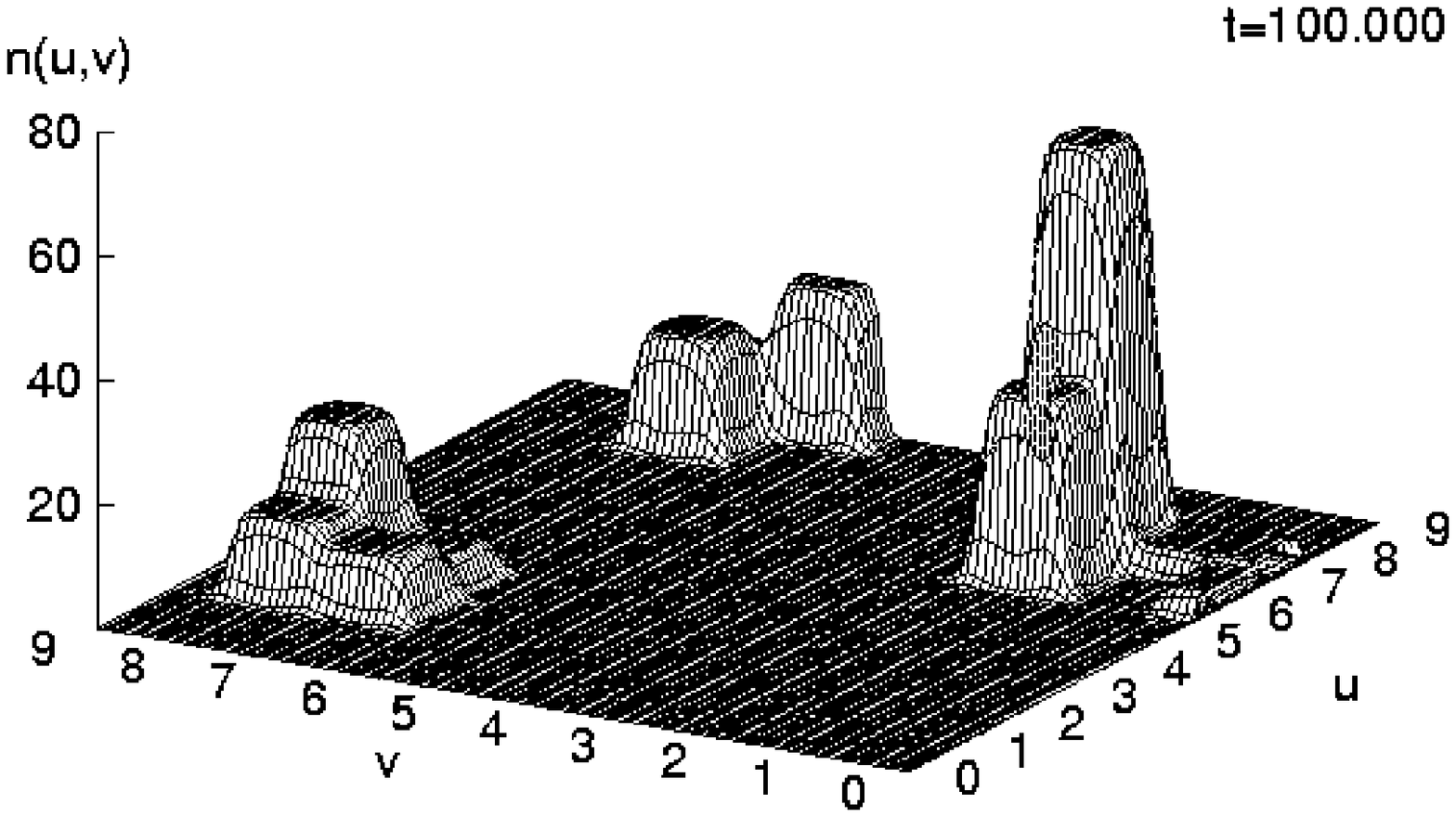,width=8.8cm,angle=-90}}
\caption[fig4]{%
Spatial density of unemployed agents, $n(u,v)$, for different times $t$
(for the parameters and initial conditions of the simulation, see text).}
\end{figure}

\begin{figure}[htbp]
\label{fig5}
\centerline{\psfig{figure=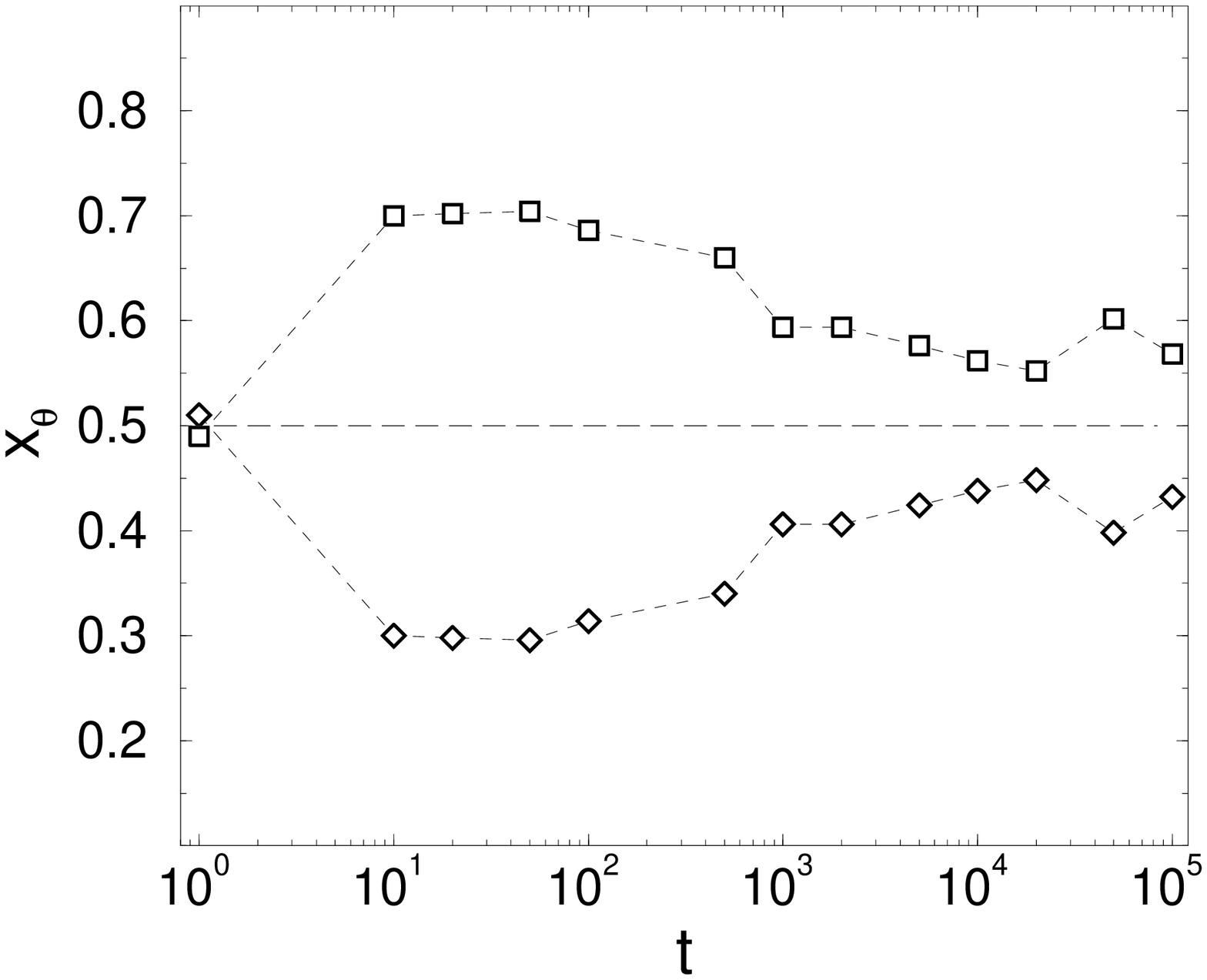,width=8cm}}
\caption[fig5]{%
  Total share $x_{\theta}$ of employed agents ($\Box$) and unemployed
  agents ($\Diamond$) vs. time $t$. The marks indicate simulation data
  which correspond to Figs. 3, 4.}
\end{figure}

The simulation shows that the evolution of the densities occurs on two
stages:

\textbf{(i)} During the first stage (for $t<1.000$) we have a significant
higher share of employed agents (up to 70 percent of the agents
community). They are broadly distributed in numerous small economic
centers, which have an employment density of about 5. This can be
understood using Figs. 1, 2. The increase of the output due to cooperative
effects allows the fast establishment of numerous small 
firms, which count on the mutual stimulations among their workers and in
the beginning offer higher wages (compared to the model without
cooperative effects). 

This growth strategy, however, is not sufficient
for output on larger scales, because the marginal product drastically
decreases (before it may increase again for an above critical employment
density).  As a result the wages may fall (possibly below the minimum
wage rate, $\omega^{\star}$), which prevents further growth. So, the
first stage is characterized by the coexistence of numerous small
economic centers.

\textbf{(ii)} During the second and much longer stage (for $t>1.000$),
some of these small centers overcome this economic bottleneck, which
allows them to grow further. In the simulations, this crossover occured
between $t=500$ and $t=800$. In the model considered here, we have only
one variable input, labor, so the crossover is mainly due to stochastic
fluctuations in $l(r,t)$. 

After the bottleneck, the marginal output
increases again, and also the wage increases with
the density of employment (cf. Figs. 1, 2). This in turn effects the
migration of the unemployed agents, which, due to \eqn{langev-red}, is
determined by two forces: the local attraction to areas with higher
wages, and the stochastic influences which keep them moving. With
$\partial \omega/\partial l >0$, the attraction may exceed the stochastic
forces, so the unemployed agents are bound to these regions, once they
got there. This, in turn is important for the further growth of these
economic centers, which needs agents to hire. As a result, we observe
on the spatial level the concentration of employed \emph{and} unemployed
agents in the \emph{same} regions. 

For their further growth, the economic centers which overcome the
bottleneck, \emph{locally} attract the labor force at the expense of the
former small economic centers. As a result of the competition process,
these small centers, which previously employed about 70 percent of the
agents, dissapear. In the new (larger) centers however, only 60 percent
of the agents can be employed (cf. Fig. 5), so the competition and the
resulting large-scale production effectively results also in an
\emph{increase of unemployment}. But the employment rate of 60 percent
seems to be a stable value for the given parameter setup, since it is
kept also in the long run, with certain fluctuations.

The concentration process described above, leads to the existence of
different \emph{extended major economic regions}, shown in Figs. 3, 4, 
which each consist of some subregions (in terms of boxes).  In the long
run (up to t=100.000) we find some remarkable features of the dynamics:
\begin{itemize}
\item[(a)] a \emph{stable coexistence} of the major economic regions,
  which keep a certain distance from each other. This \emph{critical
    distance} - which is a \emph{self-organized phenomenon} - prevents
  the regions from detracting each other. In fact, finally there is no
  force between these regions which would pull off the employed or
  unemployed agents and guide them over long distances to other regions.
  Thus, the coexistence is really a stable one.
\item[(b)] a \emph{quasi-stationary non-equilibrium} within the major
  economic centers. As we see, even in the long run, the local densities
  of employed and unemployed agents do not reach a fixed stationary
  value.  Within the major regions, there are still exchange processes
  between the participating boxes, hiring and firing, attraction and
  migration.  Hence, the total share of employed agents continues to 
  fluctuate around the mean value of 60 percent (cf. Fig. 5).
\end{itemize}

\section{Conclusions}
In this paper, we have proposed a simple dynamic model to descibe
migration and economic aggregation within the framework of Active
Brownian particles. We consider two types of particles (or agents);
employed agents, which are immobile, but (as the result of their work)
generate a wage field, and the unemployed agents which migrate by
responding to local gradients in the wage field. Further a transition
between employed and unemployed agents (hiring and firing) is considered.

The economic assumptions to be used in the model, are concluded in only
three functions: (a) the derivative $\partial \omega /\partial l$, which
describes how the local wage depends on the employment density, (b)
$k^{+}$, (c) $k^{-}$, which are the hiring and firing rates. These
functions can be determined using a production function, $Y(l)$, which
describes the output of production dependent on the variable input of
labor. Our ansatz for $Y(l)$ specifically counts on the influence of
cooperative interactions between the agents on a certain scale of
production. 

As the result of our dynamic model, we find the establishment of distinct
economic centers out of a random initial distribution. The evolution of
these centers occurs in two different stages. During the first stage,
small economic centers are formed based on the positive feedback of
mutual stimulation/cooperation among the agents. During the second and
much longer stage, some of these small centers grow at the expense of
others, which leads to the interesting situation that in our model
economic growth and decline occur \emph{at the same time}, but at
different locations, which results in specific \emph{spatial-temporal}
patterns. The competition finally leads to the concentration of the labor
force in different extended economic regions. This crossover to
large-scale production is accompanied by an increase in the unemployment
rate.

Although the extended economic regions are in an internal,
quasistationary non-equilibrium, we observe the stable coexistence
between these regions. This is an important result, which agrees e.g.
with the central place theory of economics \cite{Christaller, 1933,
  L\"osch, 1940}. Different from an attempt by
Krugman (1992), who also focused on this problem, we find that (i) the
coexistence is stable, even at the presence of fluctuations, (ii) the
centers not necessarily have to have the same number of employed agents,
to coexist, (iii) the dynamics does not simply converges into an
equilibrium state, but the centers exist in an quasistationary
non-equilibrium state and still follow a stochastic eigendynamics.

One question, which will be tackled in a forthcoming paper, is about the
influence of the different parameters. We have seen in the simulations,
that cooperative effects are able to initiate an increase in the economic
output, which leads to the establishment of many small firms. But the
final situation does not include any of those small firms and seems to be
independent of this intermediate state. This is mainly due to the fact,
that we considered only \emph{one} product. A more complex production
function with \emph{different} outputs, however, should also result in
the coexistence of small (innovative) firms and large scale production.

Another important issue is to understand the relation between the
response of the agents to local gradients in the wage field, on one hand,
and stochastic influences, on the other hand. Most likely, there exist
critical parameters which may describe a ``phase
transition'' within the agents behavior, which eventually results in
different types of spatial-temporal coexistence.

Finally, one may argue that the simple model proposed here, is not the
simplest one to describe the economic aggregation. Hence, the
minimalistic agents discussed in Sect. 1, could be even more
minimalistic. One example is the response of the employed agents to
gradients in the wage field, expressed in the parameter $q$ which enters
the firing rate, $k^{-}$.  $q=0$ simply means that internal reasons are
neglected, and the agent is fired only due to external reasons. This
might be also a reasonable assumption, however, it includes a pitfall. In
a growing economy, eventually almost all unemployed agents might be hired
at some locations. But, as long as $dY/dl > \omega^{\star}$, employed
agents are almost never fired, thus there is a shortage of free agents
for further growth. Eventually the dynamics sticks in a dead-lock
situation, because agents cannot move even if there are locations in
their neighborhood which may offer them a higher wage.  Therefore, it is
reasonable to choose $q>0$, but it is important to understand the
role of $q$, for instance, for the final patterns observed. 

So, we conclude that our model may serve as a toy model for simulating
the influence of different social and economic assumptions on the
spatial-temporal patterns in migration and economic aggregation. 

\section*{Acknowledgements}
This work has been supported by the Deutsche Forschungsgemeinschaft
(Bonn, Germany). I would like to thank J. Lobo for discussions on an
early draft of this paper, and W. Ebeling, L. Schimansky-Geier and G.
Silverberg for comments on the final version.

\section*{References}
\begin{description}
\setlength{\itemsep}{-0pt}
\setlength{\parsep}{-0pt}
\bib{anded-88}{Anderson, P.W., Arrow, K.J., Pines, D. (eds.)}{\emph{The
  Economy as an Evolving Complex System}, Reading, MA: Addison
  Wesley}{1988}

\bib{arthur-93}{Arthur, W. B.}{On Designing Economic Agents that Behave
  Like Human Agents, \emph{Journal of Evolutionary Economics} \textbf{3},
  1--22}{1993}

\bib{arthur-et-97}{Arthur, W. B.; Durlauf, S. N.; Lane, D.
  (eds.)}{\emph{The Economy as an Evolving Complex System II}, Reading,
  MA: Addison Wesley}{1997}


\bib{ben-jacob-et-95}{Ben-Jacob, E.; Cohen, I., Czir\'ok, A.}{Smart
  bacterial colonies: From complex patterns to cooperative evolution,
  \emph{Fractals} \textbf{3}, 849-868}{1995}

\bib{case-fair-92}{Case, K.E.; Fair, R.C.}{\emph{Principles of
    Economics}, Englewood Cliffs: Prentice-Hall}{1992}

\bib{christaller-33}{Christaller, W.}{\emph{Central Places in Southern
    Germany}, Jena: Fischer (English translation by C.W. Baskin, London:
  Prentice Hall, 1966)}{1933}

\bib{dendrinos-haag-84}{Dendrinos, D.; Haag, G.}{Towards a stochastic
  theory of location: Empirical evidence, \emph{Geogr. Anal.}
  \textbf{16}, 287-300}{1984}

\bib{eb-fs-tilch-98}{Ebeling, W.; Schweitzer, F.; Tilch, B.}{Active
  Brownian Particles with Energy Depots Modelling Animal Mobility,
  \emph{BioSystems} (in press)}{1998}

\bib{epstein-axtell-96}{Epstein, J. M.; Axtell, R.}{\emph{Growing
    Artificial Societies: Social Science from the Bottom Up}, Cambridge,
  MA: MIT Press/Brookings}{1996}

\bib{feisteb-89}{Feistel, R., Ebeling, W.}{\emph{Evolution of Complex
    Systems.  Self-Organization, Entropy and Development}, Dordrecht:
  Kluwer}{1989}


\bib{fujita-89}{Fujita, M.}{\emph{Urban Economic Theory}, Cambridge:
  Cambridge University Press}{1989}

\bib{haag-94}{Haag, G.}{The Rank-Size Distribution of Settlements as a
  Dynamic Multifractal Phenomenon, \emph{Chaos, Solitons \& Fractals}
  \textbf{4}, 519-534}{1994}
  
\bib{haag-munz-pumain-sanders-92}{Haag, G.; Munz, M.; Pumain, P.;
  Sanders, L.; Saint-Julien, Th.}{Interurban migration and the dynamics
  of a system of cities: 1. The stochastic framework with an application
  to the French urban system, \emph{Environment and Planning A}
  \textbf{24}, 181-198}{1992}

\bib{helb-schw-et-97}{Helbing, D.; Schweitzer, F.; Keltsch, J.;
  Moln\'{a}r, P.}{Active Walker Model for the Formation of Human and
  Animal Trail Systems, \emph{Physical Review E} \textbf{56/3},
  2527-2539}{1997}

\bib{henderson-88}{Henderson, J.V.}{\emph{Urban Development. Theory,
    Fact, and Illusion}, Oxford: Oxford University press}{1988}

\bib{holland-miller-91}{Holland, J.; Miller, J.}{Adaptive Agents in
  Economic Theory, \emph{American Economic Review Papers and Proceedings}
  \textbf{81}, 365--370}{1991}

\bib{kirman-93}{Kirman, A.}{Ants, Rationality, and Recruitment, \emph{The
    Quarterly Journal of Economics}, \textbf{108}, 37-155}{1993}


\bib{krugm-91}{Krugman, P.}{Increasing returns and economic geography,
  \emph{Journal of Political Economy} \textbf{99}, 483--499}{1991}

\bib{krugm-92b}{Krugman, P.}{A dynamic spatial model, \emph{National
    Bureau of Economic Resarch Working Paper} No. 4219}{1992}

\bib{krugm-96a}{Krugman, P.}{Urban Concentration: The role of increasing
  returns and transportation costs, \emph{Intern. Regional Science
    Review} \textbf{19}, 5-30}{1996 a}

\bib{krugm-96b}{Krugman, P.}{\emph{The Self-Organizing  Economy}, Oxford:
  Blackwell}{1996 b}

\bib{lam-95}{Lam, L.}{Active Walker Models for Complex Systems,
  \emph{Chaos, Solitons \& Fractals} \textbf{6}, 267-285}{1995}

\bib{lam-po-93}{Lam, L.; Pochy, R.}{Active-Walker Models: Growth and Form
  in Nonequilibrium Systems, \emph{Computers in Physics} \textbf{7},
  534-541}{1993}

\bib{lane-92}{Lane, D.}{Artificial Worlds and Economics, \emph{Journal of
    Evolutionary Economics} \textbf{3}, 89--107}{1992}


\bib{loesch-40}{L\"osch, A.}{\emph{The Economics of Location}, Jena:
  Fischer (English translation, New Haven: Yale University Press,
  1954)}{1940}

\bib{maesed-90}{Maes, P. (ed.)}{\emph{Designing Autonomous Agents. Theory
    and Practice From Biology to Engineering and Back}, Cambridge, MA:
  MIT Press}{1990}

\bib{marimon-et-90}{Marimon, R.; McGrattan, E.; Sargent, T. J.}{Money as
  a Medium of Exchange in an Economy with Artificially Intelligent
  Agents, \emph{Journal of Economic Dynamics and Control} \textbf{14},
  329--373}{1990}

\bib{mueller-et-97}{M\"uller, J. P.; Wooldridge, M. J.; Jennings, N. R.
  (eds.)}{\emph{Intelligent agents III : agent theories, architectures,
    and languages}, Berlin: Springer}{1997}

\bib{puu-93}{Puu, T.}{Pattern formation in spatial economics,
  \emph{Chaos, Solitons \& Fractals} \textbf{3}, 99-129}{1993}

\bib{lsg-mieth-rose-malch-95}{Schimansky-Geier, L.; Mieth, M.; Rose, H.;
  Malchow, H.}{Structure Formation by Active Brownian Particles,
  \emph{Physics Letters A} \textbf{207}, 140}{1995}

\bib{lsg-schw-mieth-97}{Schimansky-Geier, L.; Schweitzer, F.; Mieth,
  M.}{Interactive Structure Formation with Brownian Particles, \emph{in:}
  F.  Schweitzer (ed.): {\em Self-Organization of Complex Structures:
    From Individual to Collective Dynamics}, London: Gordon and Breach,
  pp.  101-118}{1997}

\bib{schw-agent-97}{Schweitzer, F.}{Active Brownian Particles: Artificial
  Agents in Physics, \emph{in:} L. Schimansky-Geier, T. P{\"o}schel
  (eds.): \emph{Stochastic Dynamics}, Berlin: Springer, pp.
  358-371}{1997}


\bib{schwei-lao-fam-97}{Schweitzer, F.; Lao, K.; Family, F.}{Active
  Random Walkers Simulate Trunk Trail Formation by Ants,
  \emph{BioSystems} \textbf{41}, 153-166}{1997}

\bib{schwei-lsg-94}{Schweitzer, F.; Schimansky-Geier, L.}{Clustering of
  Active Walkers in a Two-Component System, \emph{Physica A} {\bf 206},
  359-379}{1994}

\bib{fs-silverb-98-ed}{Schweitzer, F.; Silverberg, G.
  (eds.)}{\emph{Evolution and Self-Organization in Economics}, Berlin:
  Duncker \& Humblot}{1998}

\bib{schw-steinb-97}{Schweitzer, F.; Steinbrink, J.}{Urban Cluster Growth:
  Analysis and Computer Simulation of Urban Aggregations, in: F.
  Schweitzer (ed.), {\em Self-Organization of Complex Structures: From
    Individual to Collective Dynamics}, London: Gordon and Breach, pp.
  501-518}{1997}

\bib{silverb-97}{Silverberg, G.}{Is there Evolution after Economics?,
  \emph{in:} F. Schweitzer (ed.): {\em Self-Organization of Complex
    Structures: From Individual to Collective Dynamics}, London: Gordon
  and Breach, pp.  415-425}{1997}

\bib{silverb-versp-94}{Silverberg, G.; Verspagen, B.}{Collective
  Learning, Innovation and Growth in a Boundedly Rational, Evolutionary
  World, \emph{Journal of Evolutionary Economics} \textbf{4},
  207-226}{1994}

\bib{steels-95}{Steels, L. (ed.)}{\emph{The biology and technology of
    intelligent autonomous agents}, Berlin: Springer}{1995}


\bib{steuern-eb-94}{Steuernagel, O., Ebeling, W., Calenbuhr, V.}{An
    Elementary Model for Directed Active Motion, \emph{Chaos, Solitons \&
    Fractals} {\bf 4}, 1917-1930}{1994}
  
\bib{weibin-91}{Wei-Bin, Z.}{Synergetic Economics, New York:
  Springer}{1991}

 \bib{weidl-91}{Weidlich, W.}{Physics and Social Science -- The Approach of
  Synergetics, \emph{Physics Reports} {\bf 204}, 1-163}{1991}


\bib{weidl-97}{Weidlich, W.}{From Fast to Slow Processes in the Evolution
  of Urban and Regional Settlement Structures, \emph{in:} F. Schweitzer
  (ed.): {\em Self-Organization of Complex Structures: From Individual to
    Collective Dynamics}, London: Gordon and Breach, pp. 475-488}{1997}


  
\bib{weidl-munz-90a}{Weidlich, W.; Munz, M.}{Settlement formation, I.  A
  dynamic theory; \emph{Annals of Regional Science} \textbf{24},
  83-106}{1990 a}
  
\bib{weidl-munz-90b}{Weidlich, W.; Munz, M.}{Settlement formation, II.
  Numerical simulation; \emph{Annals of Regional Science} \textbf{24},
  177-196}{1990 b}

\end{description}

\end{document}